\documentclass[aps,floatfix,preprint,preprintnumbers,nofootinbib,superscriptaddress,natbib]{revtex4}

\usepackage{graphicx,float}
\usepackage{epsfig}		
\usepackage{dcolumn}
\usepackage{bm}
\usepackage{subfigure}

\usepackage[all]{xy}
\usepackage{amsmath,upgreek}
\usepackage{amssymb}

\usepackage{pdfpages}
\usepackage{color}
\usepackage{graphicx,epstopdf}
\usepackage[colorlinks=true,
 linkcolor=red,
 urlcolor=purple,
 citecolor=blue,hyperindex]{hyperref}
\def\0{\mbox{\tiny $0$}}
\def\1{\mbox{\tiny $1$}}
\def\2{\mbox{\tiny $2$}}
\def\3{\mbox{\tiny $3$}}
\def\4{\mbox{\tiny $4$}}
\def\5{\mbox{\tiny $5$}}
\def\6{\mbox{\tiny $6$}}
\def\7{\mbox{\tiny $7$}}
\def\8{\mbox{\tiny $8$}}
\def\9{\mbox{\tiny $9$}}

\def\f14{\mbox{\tiny $\frac{1}{4}$}}

\begin{document}

\title{Generalized phase-space description of non-linear Hamiltonian systems and the Harper-like dynamics}

\renewcommand{\baselinestretch}{1.2}
\author{A. E. Bernardini}
\email{alexeb@ufscar.br}
\altaffiliation[On leave of absence from]{~Departamento de F\'{\i}sica, Universidade Federal de S\~ao Carlos, PO Box 676, 13565-905, S\~ao Carlos, SP, Brasil.}
\author{O. Bertolami}
\email{orfeu.bertolami@fc.up.pt}
\altaffiliation[Also at~]{Centro de F\'isica das Universidades do Minho e do Porto, Rua do Campo Alegre s/n, 4169-007, Porto, Portugal.} 
\affiliation{Departamento de F\'isica e Astronomia, Faculdade de Ci\^{e}ncias da
Universidade do Porto, Rua do Campo Alegre 687, 4169-007, Porto, Portugal.}
\date{\today}

\begin{abstract}
Phase-space features of the Wigner flow for generic one-dimensional systems with a Hamiltonian, $H^{W}(q,\,p)$, constrained by the $\partial ^2 H^{W} / \partial q \partial p = 0$ condition are analytically obtained in terms of Wigner functions and Wigner currents.
Liouvillian and stationary profiles are identified for thermodynamic (TD) and Gaussian quantum ensembles to account for exact corrections due to quantum modifications over a classical phase-space pattern.
General results are then specialized to the Harper Hamiltonian system which, besides working as a feasible test platform for the framework here introduced, admits a statistical description in terms of TD and Gaussian ensembles, where the Wigner flow properties are all obtained through analytical tools.
Quantum fluctuations over the classical regime are therefore quantified through probability and information fluxes whenever the classical Hamiltonian background is provided.
Besides allowing for a broad range of theoretical applications, our results suggest that such a generalized Wigner approach works as a probe for quantumness and classicality of Harper-like systems, in a framework which can be extended to any quantum system described by Hamiltonians in the form of $H^{W}(q,\,p) = K(p) + V(q)$.
\end{abstract}


\date{\today}
\maketitle

\section{Introduction}

The Weyl-Wigner (WW) \cite{Wigner,Hillery,Ballentine,Case} phase-space formalism of quantum mechanics (QM) encompasses the dynamics of quantum systems and offers not only an enlarged but also an equivalent description of QM in terms of quasiprobability distribution functions of position and momentum coordinates. It provides subtle insights into the boundaries between quantum and classical physics as well as a straightforward access to quantum information issues \cite{Neumann,Zurek01,Zurek02,Steuernagel3}.
Without affecting the predictive power of QM, the WW formalism can also be thought of as the bridge between operator methods and path integral techniques \cite{Abr65,Sch81,Par88} encoded by a Weyl transform operation over a quantum operator $\hat{O}$, which is defined by
\begin{equation}
O^W(q,\, p)\label{Wigner111}
= 2\hspace{-.15cm} \int^{+\infty}_{-\infty} \hspace{-.35cm}ds\,\exp{\left[2\,i \,p\, s/\hbar\right]}\,\langle q - s | \hat{O} | q + s \rangle=2\hspace{-.15cm} \int^{+\infty}_{-\infty} \hspace{-.35cm} dr \,\exp{\left[-2\, i \,q\, r/\hbar\right]}\,\langle p - r | \hat{O} | p + r\rangle.
\end{equation} 
For $\hat{O}$ identified as a density matrix operator, $\hat{\rho} = |\psi \rangle \langle \psi |$, the Weyl transformed operator, $O^W(q,\, p)$, results in the so-called Wigner quasiprobability distribution function,
\begin{equation}
 h^{-1} \hat{\rho} \to W(q,\, p) = (\pi\hbar)^{-1} 
\int^{+\infty}_{-\infty} \hspace{-.35cm}ds\,\exp{\left[2\, i \, p \,s/\hbar\right]}\,
\psi(q - s)\,\psi^{\ast}(q + s),\label{Wigner222}
\end{equation}
which can also be identified as the Fourier transform of the off-diagonal elements of the associated density matrix, $\hat{\rho}$, where $h = 2\pi \hbar$ is the Planck constant. As expected, it presumes a consistent probability distribution interpretation constrained by the normalization condition over $\hat{\rho}$, that is $Tr_{\{q,p\}}[\hat{\rho}]=1$.

Notwithstanding its major role in quantum optics, which can be extended to the context of plasma and nuclear physics \cite{Sch,Lvovsky2009}, the WW phase-space framework is conveniently considered in the analysis of scattering and decoherence effects in solid state physics, for instance, in the context of semiconductor transport process phenomenology \cite{Zachos2005,Jung09}.
In addition, the formalism also provides interesting perspectives for
the interpretation of the wave-function collapse \cite{Zurek02,Bernardini13A,Leal2019}, for the paradigmatic understanding of standard QM \cite{Catarina,Bernardini13B,PhysicaA}, which includes a generalized correspondence between uncertainty relations and quantum observables \cite{Catarina001,Stein,Bernardini13C,Bernardini13E}, and even for the investigation of more general scenarios, from quantum chaos \cite{Chaos} to quantum cosmology \cite{JCAP18}.

From a strictly theoretical perspective \cite{Steuernagel3,NossoPaper,Meu2018}, the WW formalism admits a fluid equivalence of the phase-phase information flow, which encodes all the information provided by a quantum density matrix operator. 
Starting with the density matrix operator $W(q,\, p)$, one can map the quantum phase-space ensemble dynamics, and its associated Wigner currents, to describe how quantum fluctuations quantitatively affect probability distributions and information flows \cite{NossoPaper,Steuernagel3}.
By construction, it results in a continuity equation which is reduced to the Liouville equation in the classical domain. Quantum fluctuations coupled to non-linear effects driven by higher-order derivatives of the quantum mechanical potential introduce distortions (expressed in powers of $h$) over the Liouvillian flow, which set the boundaries between classical and quantum regimes \cite{NossoPaper,JCAP18,Meu2018,Bernardini2020-02}\footnote{Of course, it is well-known that the Wigner function also exhibits non-classical patterns which return negative values to the corresponding distribution function.
Parallel frameworks which include Husimi $Q$ \cite{Husimi,Ballentine} and Glauber-Sudarshan representations \cite{Glauber,Sudarshan,Carmichael,Callaway}, or even the more specific optical tomographic probability representation \cite{Amosov,Radon,Mancini} of QM, are used to be considered in an adequate interpretation of the negative quasiprobability distributions.}.
Another relevant aspect involving the classical-to-quantum correspondence is related to the phase-space probability and information flow interpretations devised by Wigner \cite{Wigner}, which establish the correspondence between classical mechanics and classical thermodynamics (TD) on the same level as that between QM and TD. That is, the Maxwell-Boltzmann distribution of classical statistical mechanics $\exp(- H^{W}(q,\,p)/k_B\mathcal{T})$, where $H$ is the Hamiltonian, $k_B$ is the Boltzmann constant, and $\mathcal{T}$ is the temperature, can be statistically treated on equal footing as the quantum version described by $\sum_{n}\exp(- E_n/k_B\mathcal{T}) W_n(q,\, p)$, where $n$ is a quantum number index for {\em eigen}energies, $E_n$, and Wigner-related {\em eigen}functions, $W_n(q,\, p)$.

Given the universal features of the WW formalism described above, our proposal in this work is to extend the formulation of the QM in phase-space to non-linear Hamiltonian systems generically described by a Hamiltonian of the form, 
\begin{equation}
H^{W}(q,\,p) = K(p) + V(q),
\label{nlh}
\end{equation}
which is evidently constrained by the $\partial ^2 H^{W} / \partial q \partial p = 0$ condition, where $K(p)$ and $V(q)$ are arbitrary functions of $p$ and $q$, respectively.
The point here is that QM states are based on Hilbert spaces and operators implemented through the Schr\"odinger equation, whereas classical mechanics is geometrically defined on symplectic manifolds, whose dynamic trajectories are described by Hamilton's equations.
For more general Hamiltonians, like $H^{W}(q,\,p) = K(p) + V(q)$, implementing the {\em Hamiltonian function} through an {\em eigen}system, $H^{W}\, \psi_n = E_n\, \psi_n$ is sometimes unfeasible.
That is often the case for $K(p)$ and $V(q)$ simultaneously described by non-polynomial functions of $p$ and $q$, respectively, for which the corresponding Schr\"odinger-like {\em eigen}system cannot be solved, even numerically. 

Alternatively, when the dynamics is described by a {\em Hamiltonian constraint}, as opposed to a Hamiltonian function, the WW formalism can be implemented through a probability flux continuity equation, through which the fluid analogy and the analytical computability of the WW formalism provide solutions for quantum ensembles, as opposed to quantum states.
Therefore, at least for (TD and Gaussian) quantum ensembles, the classical-quantum limits of exact quantum solutions can be tested by means of (quantum) probability distributions and information quantifiers, all obtained from the WW formalism. 

Hence, our starting point is to consider the Wigner flow framework for Hamiltonian systems of the form $H^{W}(q,\,p) = K(p) + V(q)$, for which the description of probability currents, and the construction of the Hamiltonian related (quantum) information quantifiers are expected.
Once developed, the framework will be applied to Harper-type \cite{Harper,Harper02} systems, which naturally encompass non-linearity properties arising from momentum coordinate, $p$. In this case, for $K(p)$ identified by a sinusoidal form $\cos(p/p_0)$, which is therefore distinct from the standard Schr\"odinger (quadratic momentum) formulation, it will be possible to confront classical and quantum scenarios in the mentioned Hamiltonian context. 

The so-called Harper Hamiltonian originally described the effect of a uniform magnetic field on a conduction-band metal \cite{Harper}, where a tight-binding approximation for symmetric cubic crystals was assumed. 
The Harper-Hofstadter extension of the formalism \cite{Harper,NatHarper,RMPRMP} was generalized to lattice systems in the presence of a gauge field, where ground-level chiral many-body states emerge from the system populated with interacting particles \cite{NatHarper}.
In the past few years, the Harper Hamiltonian has been experimentally implemented for neutral particles in optical lattices through a laser-assisted tunneling mechanism, which effectively probes the expansion of the atoms in the lattice, where a potential energy gradient is provided either by gravity or by magnetic fields \cite{PRL-Harper}.
The laser-assisted tunneling processes introduce single site and single particle resolved controls which engender an image platform that enables one to build chiral systems and then to tune the particle number within a chiral state, atom by atom.
This allows for controlling the lattice size \cite{PRL-Harper,PRA-Harper19} exhibiting the band structure displayed by the so-called Hofstadter's butterfly \cite{Harper,Harper02}.
The semiclassical interpretation of the original tight-binding approach for the Harper model results in a classical effective one-dimensional Hamiltonian that reads 
\begin{equation}\label{HamHarper01}
H^{W}_H(q,\, p)= \pm \left[2\pi\alpha\cos(p/p_0) + 2\pi\alpha\cos(q/q_0)\right],
\end{equation}
for an isotropic $q-p$ phase-space where the associated operators, $\hat{q}$ and $\hat{p}$ are driven by QM-like (non-)commuting relations, $[\hat{q},\hat{p}] = i\,2\pi\alpha\,p_0\,q_0$, with $\alpha$ identified as the Peierls phase \cite{Harper,Harper02,PRA-Harper19}, which works as an effective modulation of the reduced Planck constant $\hbar$, in a periodic closed phase-space trajectory.
Recently, the elementary properties exhibited by the Harper-Hamiltonian, Eq. \eqref{HamHarper01}, have also motivated the research on the transition from order to chaos, through a kind of {\em kicked} time-dependent Harper map \cite{Shape,Artuso01,Artuso02,PREPRE}.
Concerns regarding the quantum topological issues of Harper-type models, Floquet systems exhibiting the Hall effect due to electromagnetic fields \cite{PRBPRB}, and the extended Aubry-Andr\'e-Harper formulation for classifying topological states \cite{PRBPRB2} have also been investigated.
Besides simulation and experimental feasibilities, Harper-like systems driven by the classical correspondence with $H_H(q,\, p)$ from (\ref{HamHarper01}) fill the conditions for $H^{W}(q,\,p) = K(p) + V(q)$ introduced above.
Therefore, once the prescribed phase-space description of non-linear ($q$ and $p$) Hamiltonian systems is established, the Harper platform might be relevant, as a test, in describing the interplay between microscopic-quantum and macroscopic-classical realities.

Given the above assumptions, the outline of this paper is as follows.
Sec.~II is concerned with the foundations of the WW framework in order to derive stationarity and quantifying Liouvillian properties related to the extended Wigner framework for $H^{W}(q,\,p) = K(p) + V(q)$ Hamiltonians.
In Sec.~III, the results are specialized to TD ensembles, which admit a precise interpretation of stationarity conditions in the context of the classical-to-quantum correspondence and are physically appealing in relation to statistical mechanics.
The same analysis is extended to Gaussian ensembles in Sec.~IV, in a Wigner flow framework in which the overall quantum fluctuations over a classical phase-space trajectory can be computed in terms of a convergent infinite series expansion over $\hbar^{2n}$. 
Through exact analytical results, the Gaussian framework accounts for complete quantum corrections of the Hamiltonian classical profile, providing a natural approach to establish accurate correspondence between classical and quantum scenarios through the information quantum correlation quantifiers obtained from Wigner currents.
In Sec.~V, the Harper Hamiltonian is considered as a test platform for the derived extended phase-space framework.
Stationarity and Liouvillianity conditions as well as parameters for classical-to-quantum correspondence are then specialized to TD and Gaussian ensembles in which the Wigner flow is driven by the Harper Hamiltonian.
These are all explicitly obtained in terms of analytical expressions for the Wigner currents for both TD and Gaussian ensembles.
Our conclusions and the outlook for further research are presented in Sec.~VI.

\section{Stationarity and Liouvillianity in the extended Wigner framework}

Based on the symmetries of the Heisenberg-Weyl group of translations, the Wigner phase-space quasidistribution function associated with a density operator $\hat{\rho}$, in the form of an overlap integral, Eq.~\eqref{Wigner222}, was proposed by Wigner's seminal work \cite{Wigner} when accounting for quantum corrections to TD equilibrium states.
The Wigner function's most elementary property is concerned with its marginal distributions, which return position and momentum distributions upon integrations over the momentum and position coordinates, respectively,
\begin{equation}
\vert \psi(q)\vert^2 = \int^{+\infty}_{-\infty} \hspace{-.35cm}dp\,W(q,\, p)
\qquad
\leftrightarrow
\qquad
\vert \varphi(p)\vert^2 = \int^{+\infty}_{-\infty} \hspace{-.35cm}dq\,W(q,\, p),
\end{equation}
such that the Fourier transform of the respective wave functions,
\begin{equation}
 \varphi(p)=
(2\pi\hbar)^{-1/2}\int^{+\infty}_{-\infty} \hspace{-.35cm} dq\,\exp{\left[i \, p \,q/\hbar\right]}\,
\psi(q),
\end{equation}
is the property, intrinsically based on the Hilbert space features of the Schr\"odinger QM, that suppresses the coexistence of positive-definite position and/or momentum probability distributions.

The connection of the Wigner function to the matrix operator QM (cf. Eqs.~\eqref{Wigner111} and \eqref{Wigner222}) allows for computing the averaged values of quantum observables described by generic operators, $\hat{O}$, evaluated through an overlap integral over the infinite volume described by the phase-space coordinates, $q$ and $p$, as \cite{Wigner,Case}
\begin{equation}
 \langle O \rangle = 
\int^{+\infty}_{-\infty} \hspace{-.35cm}dp\int^{+\infty}_{-\infty} \hspace{-.35cm} {dq}\,\,W(q,\, p)\,{O^W}(q,\, p), \label{eqfive}
\end{equation}
which corresponds to the trace of the product between $\hat{\rho}$ and $\hat{O}$, $Tr_{\{q,p\}}\left[\hat{\rho}\hat{O}\right]$.
Moreover, the statistical aspects inherent to the $W(q,\, p)$ definition also admit extensions from pure states to statistical mixtures, through which, for instance, the replacement of ${O^W}(q,\, p)$ by $W(q,\, p)$ into Eq.~\eqref{eqfive} leads to the quantum purity computed through an analogous of the trace operation, $Tr_{\{q,p\}}[\hat{\rho}^2]$, read as
\begin{equation}
Tr_{\{q,p\}}[\hat{\rho}^2] = 2\pi\hbar\int^{+\infty}_{-\infty}\hspace{-.35cm}dp\int^{+\infty}_{-\infty} \hspace{-.35cm} {dq}\,\,\,W(q,\, p)^2,
\label{eqpureza}
\end{equation}
satisfying the pure state constraint, $Tr_{\{q,p\}}[\hat{\rho}^2] = Tr_{\{q,p\}}[\hat{\rho}] = 1$.

In addition, flow properties of the Wigner function, $W(q,\,p) \to W(q,\,p;\,t)$, are also connected to the Hamiltonian dynamics.
These properties are described by a vector flux \cite{Steuernagel3,NossoPaper,Meu2018}, $\mathbf{J}(q,\,p;\,t)$, decomposed into the phase-space coordinate directions, $\hat{q}$ and $\hat{p}$, as $\mathbf{J} = J_q\,\hat{q} + J_p\,\hat{p}$, in order to 
reproduce a flow field connected to the Wigner function dynamics through the continuity equation \cite{Case,Ballentine,Steuernagel3,NossoPaper,Meu2018},
\begin{equation}
{\partial_t W} + {\partial_q J_q}+{\partial_p J_p} =0,
\label{alexquaz51}
\end{equation}
where the shortened notation for partial derivatives is set as $\partial_a \equiv \partial/\partial a$. In this case, for a non-relativistic Hamiltonian operator, ${H}(\hat{Q},\,\hat{P})$, from which the Weyl transforms yields 
\begin{equation}
{H}(\hat{Q},\,\hat{P}) = \frac{\hat{P}^2}{2m} + V(\hat{Q}) \quad\to \quad H^{W}(q,\, p) = \frac{{p}^2}{2m} + V(q),
\end{equation}
one has \cite{Case,Ballentine,Steuernagel3,NossoPaper}
\begin{equation}
J_q(q,\,p;\,t)= \frac{p}{m}\,W(q,\,p;\,t), \label{alexquaz500BB}
\end{equation}
and
\begin{equation}
J_p(q,\,p;\,t) = -\sum_{\eta=0}^{\infty} \left(\frac{i\,\hbar}{2}\right)^{2\eta}\frac{1}{(2\eta+1)!} \, \left[\partial_q^{2\eta+1}V(q)\right]\,\partial_p ^{2\eta}W(q,\,p;\,t),
\label{alexquaz500}
\end{equation}
with $\partial^s_a \equiv (\partial/\partial a)^s$, from which one notices that the above identified series expansion contributions from $\eta \geq 1$ introduce the quantum corrections which distort classical trajectories. In fact, for Eq.~\eqref{alexquaz500}, the suppression of the $\eta \geq 1$ contributions results in a classical Hamiltonian description of the phase-space probability distribution dynamics in terms of classical Wigner currents,
\begin{equation}
J^{\mathcal{C}}_q(q,\,p;\,t)= +({\partial_p H^{W}})\,W(q,\,p;\,t), \label{alexquaz500BB2}
\end{equation}
and
\begin{equation}
J^{\mathcal{C}}_p(q,\,p;\,t) = -({\partial_q H^{W}})\,W(q,\,p;\,t),
\label{alexquaz500CC2}
\end{equation}
which, once substituted into Eq.~\eqref{alexquaz51}, return the (classical) Liouville equation, with the classical phase-space velocity identified by $\mathbf{v}_{\xi(\mathcal{C})} = \dot{\mbox{\boldmath $\xi$}} = (\dot{q},\,\dot{p})\equiv ({\partial_p H^{W}},\,-{\partial_q H^{W}})$, with $\mbox{\boldmath $\nabla$}_{\xi}\cdot \mathbf{v}_{\xi(\mathcal{C})}= \partial_q \dot{q} + \partial_p\dot{p} = 0$, where {\em dots} denote the time derivative, $d/dt$.
Likewise, for a quantum current parameterized by $\mathbf{J} = \mathbf{w}\,W$, where the Wigner phase-space velocity, $\mathbf{w}$, is the quantum analog of $\mathbf{v}_{\xi(\mathcal{C})}$, a suitable divergent behavior is identified by
\begin{equation}
\mbox{\boldmath $\nabla$}_{\xi} \cdot \mathbf{w} = \frac{W\, \mbox{\boldmath $\nabla$}_{\xi}\cdot \mathbf{J} - \mathbf{J}\cdot\mbox{\boldmath $\nabla$}_{\xi}W}{W^2},
\label{zeqnz59}
\end{equation}
since $\mbox{\boldmath $\nabla$}_{\xi}\cdot\mathbf{J} = W\,\mbox{\boldmath $\nabla$}_{\xi}\cdot\mathbf{w}+ \mathbf{w}\cdot \mbox{\boldmath $\nabla$}_{\xi}W$ \cite{Steuernagel3}.
Therefore, Wigner function stationarity and Liouvillian behaviors are straightforwardly identified and quantified in terms of Eqs.~\eqref{alexquaz51} and \eqref{zeqnz59}, by $\mbox{\boldmath $\nabla$}_{\xi} \cdot \mathbf{J} = 0$ and $\mbox{\boldmath $\nabla$}_{\xi} \cdot \mathbf{w} = 0$, respectively.

This Wigner current constraint, $\mathbf{J} = \mathbf{w}\,W$, indeed brings some relevant aspects of the topological properties of the Wigner flow. Considering the integral version of the above-mentioned stationary and Liouvillian conditions, these properties can be captured by the introduction of a {\em substantial derivative} \cite{Gradshteyn,MeuPT,NossoPaper} written as 
\begin{equation}
\frac{D~}{Dt} \int_{V}dV\,{f(W)} \equiv 
\int_{V}dV\,\left(\frac{D}{Dt} + \mbox{\boldmath $\nabla$}_{\xi}\cdot \mathbf{v}_{\xi}\right){f(W)}\label{alexquaz57D},
\end{equation}
which evinces the conservative features associated with the elementary vector unity of the phase-space, $\mbox{\boldmath $\xi$}$, surrounded by an infinitesimal volume, $dV \equiv dp\,dq$. For the case of a volume $V$ enclosed by the two-dimensional comoving closed surface, $\mathcal{C}$, it gives rise to the usual tool for quantifying the flux of information through $\mathcal{C}$ \cite{NossoPaper}, which is depicted by the trajectory obtained from the classical velocity, $\mathbf{v}_{\xi(\mathcal{C})}$, which can be seen from Eq.~\eqref{alexquaz57D}, when $f(W)$ is identified by a power of $W$, $W^{\beta}$,
\begin{eqnarray}
 \frac{D~}{Dt}\left(\int_{V_{\mathcal{C}}}dV\, {W}^{\beta}\right)
&=& \int_{V_{\mathcal{C}}}dV\,\left[\frac{D~}{Dt} ({W}^{\beta}) + {W}^{\beta} \mbox{\boldmath $\nabla$}_{\xi}\cdot \mathbf{v}_{\xi(\mathcal{C})}\right]\nonumber\\
&=&\int_{V_{\mathcal{C}}}dV\,\left[\partial_t ({W}^{\beta}) + \mbox{\boldmath $\nabla$}_{\xi}\cdot(\mathbf{v}_{\xi(\mathcal{C})} {W}^{\beta})\right]\nonumber\\
&=&\int_{V_{\mathcal{C}}}dV\,\left[-\beta{W}^{\beta-1} \mbox{\boldmath $\nabla$}_{\xi}\cdot\mbox{\boldmath${J}$} + \mbox{\boldmath $\nabla$}_{\xi}\cdot(\mathbf{v}_{\xi(\mathcal{C})} {W}^{\beta})\right]\nonumber\\
&=&-\int_{V_{\mathcal{C}}}dV\,\left[(\beta-1){W}^{\beta} \,\mbox{\boldmath $\nabla$}_{\xi}\cdot\mathbf{w} 
+ \mbox{\boldmath $\nabla$}_{\xi}\cdot(\mbox{\boldmath${J}$}{W}^{\beta-1} - \mathbf{v}_{\xi(\mathcal{C})} {W}^{\beta})\right]\nonumber\\
 &=& -\int_{V_{\mathcal{C}}}dV\,(\beta-1){W}^{\beta} \,\mbox{\boldmath $\nabla$}_{\xi}\cdot\mathbf{w} 
- \oint_{\mathcal{C}}d\ell\, {W}^{\beta-1}\left(\mbox{\boldmath${J}$}\cdot \mathbf{n}\right),\label{quaz692}
\end{eqnarray}
with the unitary vector, $\mathbf{n}$, satisfying $\mathbf{n}\cdot\mathbf{v}_{\xi(\mathcal{C})}= 0$.

For $\beta = 0$, Eq.~\eqref{quaz692} results in
\begin{eqnarray}
 \frac{D}{Dt}V
 &=& \int_{V_{\mathcal{C}}}dV \,\mbox{\boldmath $\nabla$}_{\xi}\cdot\mathbf{w} 
- \oint_{\mathcal{C}}d\ell\,\mathbf{w} \cdot \mathbf{n}= 0,\label{quaz69222}
\end{eqnarray}
which reflects the phase-space volume conservation, given that Green's theorem states that
\begin{eqnarray}
\int_{V_{\mathcal{C}}}dV \,\mbox{\boldmath $\nabla$}_{\xi}\cdot\mathbf{w} 
 &=& \oint_{\mathcal{C}}d\ell\,\mathbf{w} \cdot \mathbf{n} ,\label{quaz69222}
\end{eqnarray}
which of course vanishes for $\mbox{\boldmath $\nabla$}_{\xi} \cdot \mathbf{w} = 0$, in a kind of integral version of the Liouville condition.
Analogously, by identifying the phase-space volume integrated probability as
\begin{equation}
\sigma_{(\mathcal{C})} =\int_{V_{_{\mathcal{C}}}}dV\,{W},
\label{quaz6022}
\end{equation}
for $\beta = 1$, Eq.~\eqref{quaz692} results in
\begin{eqnarray}
 \frac{D}{Dt}\sigma_{(\mathcal{C})}
 &=& - \oint_{\mathcal{C}}d\ell\,\mathbf{J}\cdot \mathbf{n},\label{quaz69222}
\end{eqnarray}
which reflects the probability conservation for stationary states, given that Green's theorem now states that
\begin{eqnarray}
\int_{V_{\mathcal{C}}}dV \,\mbox{\boldmath $\nabla$}_{\xi}\cdot\mathbf{J} 
 &=& \oint_{\mathcal{C}}d\ell\,\mbox{\boldmath${J}$}\cdot \mathbf{n}.\label{quaz}
\end{eqnarray}

As proposed, the above set of stationarity and Liouvillianity quantifying tools shall be evaluated in the extended Wigner framework.
In this case, explicit expressions for the Wigner current components, $\mathbf{J} = J_q\,\hat{q} + J_p\,\hat{p}$, must be deduced for QM Hamiltonians generically identified by the Weyl transform \eqref{nlh}.

In what follows such an extended form of the Wigner continuity equation is obtained
by deducing the time dependence of $W$. As a guide, our departing point is the Von Neumann equation for the state density operator, $\hat{\rho} = \vert \psi \rangle \langle \psi\vert$, obtained from \cite{Ballentine}
\begin{equation}
\partial_t\hat{\rho} = i\hbar^{-1} \left[\hat{\rho}, \, H \right] \equiv{\partial_t^{^{(K)}}\hat{\rho}} ~ + ~{\partial_t^{^{(V)}}\hat{\rho}},\quad \mbox{with} ~~ {\partial_t^{^{( \mathcal{A})}}\hat{\rho}} = i\hbar^{-1} \left[\hat{\rho}, \, \mathcal{A}\right],
\label{dens}
\end{equation}
which can then be separately evaluated in momentum and position representations, for $\mathcal{A} \equiv K(\hat{P}), \,V(\hat{Q})$.
Hence, using the Wigner function properties from Eq.~\eqref{Wigner222} to transform each contribution into its respective Wigner representation (cf. Ref.~\cite{Ballentine} for non-relativistic QM):
\begin{eqnarray}
\partial^{^{(K)}}_t\langle p \vert {\rho} \vert p'\rangle &=& i \hbar^{-1}\langle p \vert {\rho} \vert p'\rangle
\,\left(K(p') - K(p)\right)\Rightarrow\\
 \partial^{^{(K)}}_t W(q,\,p;\,t) &=& i \hbar^{-1} (\pi\hbar)^{-1}
\int^{+\infty}_{-\infty} \hspace{-.35cm}dr\,
\rho^{W,\varphi}_{(p-r;\,p+r)}
\exp{\left[-2\, i \, q \,r/\hbar\right]}
\,\left[K(p+r) - K(p-r)\right],\nonumber
\label{Wigner222BB}
\end{eqnarray}
where $\rho^{W,\varphi}_{(p-r;\,p+r)} \equiv \langle p-r \vert {\rho} \vert p+r\rangle$ has been identified by $\varphi(p- r)\,\varphi^{\ast}(p+ r)$, and
\begin{eqnarray}
\partial^{^{(V)}}_t\langle q \vert {\rho} \vert q'\rangle &=& i \hbar^{-1}\langle q \vert {\rho} \vert q'\rangle
\,\left[V(q') - V(q)\right]\Rightarrow\\
 \partial^{^{(V)}}_t W(q,\,p;\,t) &=& i \hbar^{-1} (\pi\hbar)^{-1}
\int^{+\infty}_{-\infty} \hspace{-.35cm}ds\,
\rho^{W,\psi}_{(q-s;\,q+s)}
\exp{\left[2\, i \, p \,s/\hbar\right]}
\,\left(V(q+s) - V(q-s)\right),\nonumber
\label{Wigner222CC}
\end{eqnarray} 
where $\rho^{W,\psi}_{(q-s;\,q+s)} \equiv \langle q - s \vert {\rho} \vert q + s\rangle$ has been identified by $\psi(q - s)\,\psi^{\ast}(q + s)$.
Now, by noticing that
\begin{equation}
K(p+r) - K(p-r) = 2\sum_{\eta=0}^{\infty}\frac{r^{2\eta+1}}{(2\eta+1)!} \,\partial_p^{2\eta+1}K(p),
\label{alexquaz500BB}
\end{equation}
and
\begin{equation}
V(q+s) - V(q-s) = 2\sum_{\eta=0}^{\infty}\frac{s^{2\eta+1}}{(2\eta+1)!} \,\partial_q^{2\eta+1}
V(q),
\label{alexquaz500CC}
\end{equation}
and identifying the auxiliary variables, $r$ and $s$, by $+i(\hbar/2)\, \partial_q$ (cf. Eq~\eqref{Wigner222BB}) and $-i(\hbar/2)\, \partial_p$ (cf. Eq~\eqref{Wigner222CC}), respectively, one recovers an equivalent Wigner continuity equation cast in the form of Eq.~\eqref{alexquaz51}, i.e. with the stationarity quantifier explicitly given by
\begin{equation} \label{helps}
\partial_t W= \sum_{\eta=0}^{\infty}\frac{(-1)^{\eta}\hbar^{2\eta}}{2^{2\eta}(2\eta+1)!} \, \left\{
\left[\partial_q^{2\eta+1}V(q)\right]\,\partial_p^{2\eta+1}W
-
\left[\partial_p^{2\eta+1}K(p)\right]\,\partial_q^{2\eta+1}W
\right\},\end{equation}
from which one has
\begin{equation}
J_q(q,\,p;\,t) = +\sum_{\eta=0}^{\infty} \left(\frac{i\,\hbar}{2}\right)^{2\eta}\frac{1}{(2\eta+1)!} \, \left[\partial_p^{2\eta+1} K(p)\right]\,\partial_q^{2\eta}W(q,\,p;\,t),
\label{alexquaz500BB}
\end{equation}
and
\begin{equation}
J_p(q,\,p;\,t) = -\sum_{\eta=0}^{\infty} \left(\frac{i\,\hbar}{2}\right)^{2\eta}\frac{1}{(2\eta+1)!} \, \left[\partial_q^{2\eta+1} V(q)\right]\,\partial_p^{2\eta}W(q,\,p;\,t),\label{alexquaz500CC}
\end{equation}
that encompass all the contributions for quantum corrections.
Analogously, to fully capture the quantum effects over the classical Hamiltonian regime described by currents in the form of \eqref{alexquaz500BB2}-\eqref{alexquaz500CC2}, the Liouvillianity quantifier (as from Eq.~\eqref{zeqnz59}) is expressed by
\begin{equation}
\mbox{\boldmath $\nabla$}_{\xi} \cdot \mathbf{w} = \sum_{\eta=1}^{\infty}\frac{(-1)^{\eta}\hbar^{2\eta}}{2^{2\eta}(2\eta+1)!}
\left\{
\left[\partial_p^{2\eta+1}K(p)\right]\,
\partial_q\left[\frac{1}{W}\partial_q^{2\eta}W\right]
-
\left[\partial_q^{2\eta+1}V(q)\right]\,
\partial_p\left[\frac{1}{W}\partial_p^{2\eta}W\right]
\right\}. ~~~\end{equation}
Finally, as it will be pointed out in the following sections, the above presented forms for quantifiers of stationarity and Liouvillian regimes provide the fundamental information content of quantum ensembles, which is essential for discriminating classical from quantum behaviors.

\section{Thermodynamic ensembles}

The Wigner function is also an essential tool for defining quantum projectors for TD ensembles.
Once it is assumed that the spectral decomposition of the quantum system driven by the Hamiltonian \eqref{nlh} cannot be obtained, the stationary solution of the Wigner continuity equation (cf. Eq.~\eqref{alexquaz51}) for the corresponding quantum TD ensemble can be computed through a perturbative expansion over the classical Maxwell-Boltzmann distribution \cite{Wigner,Coffey07}. 

To find quantum corrections to the classical TD regime, one solves a closed equilibrium equation which indeed exhibits the response of the quantum system to the thermal equilibrium for relatively high temperatures. It yields the corresponding Wigner stationary distribution, $W_{St}(q,\,p;\,\beta)$, where $\beta = 1/ k_{B} \mathcal{T}$.
In fact, one departs from a classical distribution described by
\begin{equation}
W_0(q,\,p;\,\beta) = (\hbar\mathcal{Z}_0(\beta))^{-1}\,\exp[-\beta H^W(q,\,p)],
\end{equation}
with the partition function identified by
\begin{equation}
 \mathcal{Z}_0(\beta) = 
\hbar^{-1}\int^{+\infty}_{-\infty} \hspace{-.35cm}dq\,\int^{+\infty}_{-\infty} \hspace{-.35cm}dp\,\exp[-\beta H^W(q,\,p)],
\end{equation}
and notices that the stationary solution of Eq.~\eqref{alexquaz51} is iteratively obtained through a series expansion \cite{Wigner,Coffey07},
\begin{equation}
W^{(2N)}_{St}(q,\,p;\,\beta) = \sum_{\eta=0}^N\,\hbar^{2\eta}\,W_{2\eta}(q,\,p;\,\beta),
\end{equation}
truncated at order $\mathcal{O}(\hbar^{2N})$, from which $$\lim_{N\to \infty}W^{(2N)}_{St}(q,\,p;\,\beta) = W_{St}(q,\,p;\,\beta).$$

The assumption that the TD equilibrium prevails everywhere in the phase-space ensures that a closed equation resolution procedure for determining, order by order, the contributions from $W_{2\eta}$ applies \cite{Wigner,Coffey07}. 
Given the arbitrariness of $K(p)$ and $V(q)$, in this first analysis, the quantum corrections shall be constrained to contributions up to order $\mathcal{O}(\hbar^{2})$.
In this case, one identifies $W_{2}(q,\,p;\,\beta)$ with $\chi_{(q,\,p;\,\beta)}\, W_{0}(q,\,p;\,\beta)$, where $\chi_{(q,\,p;\,\beta)}$ is a multiplicative parameter that factorizes the corrections over $W_{0}(q,\,p;\,\beta)$, and computes the time derivative of the stationary solution for $N=1$, $W^{(2)}_{St}(q,\,p;\,\beta) = W_{0}(q,\,p;\,\beta)+\hbar^{2}\,W_{2}(q,\,p;\,\beta)$, in order to obtain
\begin{eqnarray}\label{000}
\partial_t W^{(2)}_{St} &=&\partial_t W_{0} + \hbar^{2}\,\partial_t W_{2} = \left[1+\hbar^{2}\,\chi_{(q,\,p;\,\beta)}\right]\,\partial_t W_{0} + \hbar^{2}\,W_{0}\, \partial_t\chi_{(q,\,p;\,\beta)}.
\end{eqnarray}
Up to order $\mathcal{O}(\hbar^{2})$, the contribution from $\partial_t\chi_{(q,\,p;\,\beta)}$ into the above expression can be expressed by
\begin{eqnarray}\label{001}
\partial_t\chi_{(q,\,p;\,\beta)}&=&-\{\chi,\, H\}_{PB} = \partial_q V(q)\,\partial_p \chi_{(q,\,p;\,\beta)}- \partial_p K(p)\,\partial_q \chi_{(q,\,p;\,\beta)},\end{eqnarray}
where ``PB'' stands for the {\em Poission brackets}.
Likewise, now from Eq.~\eqref{helps}, one writes 
\begin{eqnarray}\label{002}
\partial_t W_{0} = - \mbox{\boldmath $\nabla$}_{\xi}\cdot \mathbf{J} &=& \left[\partial_q V(q)\,\partial_p W_0
-
\partial_p K(p)\,\partial_q W_0\right]\nonumber\\
&&\quad - 
\frac{\hbar^{2}}{24} \, \left[
\left(\partial_q^{3}V(q)\right)\,\partial_p^{3}W_0
-
\left(\partial_p^{3}K(p)\right)\,\partial_q^{3}W_0
\right] + \mathcal{O}(\hbar^4),
\end{eqnarray}
where, due to the classical nature of $W_0$, the contribution from classical currents vanishes, i.e.
$$\partial_q V(q)\,\partial_p W_0
-
\partial_p K(p)\,\partial_q W_0 =-\{W_0,\, H\}_{PB}= 0.$$
In the next step, after identifying the auxiliary derivatives given by $\partial_q W_0 = -\beta\,W_0\, \partial_q V(q)$ and $\partial_p W_0 = -\beta W_0 \,\partial_p K(p)$, one notices that
\begin{eqnarray}
\partial^2_q W_0 &=& -\beta\, W_0 \,\partial^2_q V(q) + \beta^2 \,W_0 \left(\partial_q V(q)\right)^2,\\
\partial^2_p W_0 &=& -\beta\, W_0 \,\partial^2_p K(p) + \beta^2 \,W_0 \left(\partial_p K(p)\right)^2,
\end{eqnarray}
and that
\begin{eqnarray}
\partial^3_q W_0 &=& -\beta\, W_0 \,\partial^3_q V(q) + 3\beta^2 \,W_0 \,\left(\partial^2_q V(q)\right)\partial_q V(q) - \beta^3\, W_0\, \left(\partial_q V(q)\right)^3,\\
\partial^3_p W_0 &=& -\beta\, W_0 \,\partial^3_p K(p) + 3\beta^2 \,W_0 \,\left(\partial^2_p K(p)\right)\partial_p K(p) - \beta^3\, W_0 \,\left(\partial_p K(p)\right)^3,
\end{eqnarray}
which can all be inserted into Eq.~\eqref{002}. Then, through the insertion of Eqs.~\eqref{001} and \eqref{002} into Eq.~\eqref{000}, it yields
\begin{eqnarray}
\partial_t W^{(2)}_{St} &=&\hbar^{2}W_{0} \bigg{\{}\partial_q V(q)\,\partial_p \chi_{(q,\,p;\,\beta)}- \partial_p K(p)\,\partial_q \chi_{(q,\,p;\,\beta)}
\nonumber\\
&& \quad \quad\quad
-\frac{\beta^2}{8} \left[
\left(\partial_q^{3}V(q)\right)\,\left(\partial^2_p K(p)\right)\partial_p K(p)
-
\left(\partial_p^{3}K(p)\right)\,\left(\partial^2_q V(q)\right)\partial_q V(q)
\right]
\nonumber\\
&& \quad\quad \quad\quad\quad
+\frac{\beta^3}{24} \left[
\left(\partial_q^{3}V(q)\right)\,\left(\partial_p K(p)\right)^3
-
\left(\partial_p^{3}K(p)\right)\,\left(\partial_q V(q)\right)^3
\right]\bigg{\}} + \mathcal{O}(\hbar^4). \quad
\end{eqnarray}
Finally, from the above result, the stationarity condition over $W^{(2)}_{St}$, $\partial_t W^{(2)}_{St} =0$, results in
\begin{eqnarray}
\chi_{(q,\,p;\,\beta)} &=&
-\frac{\beta^2}{8}
\partial_q^{2}V(q)\,\partial^2_p K(p)
+\frac{\beta^3}{24}
\left[
\partial_q^{2}V(q)\,\left(\partial_p K(p)\right)^2
+
\partial_p^{2}K(p)\,\left(\partial_q V(q)\right)^2
\right],
\end{eqnarray}
which gives
\begin{eqnarray}\label{oiue}
W^{(2)}_{St}(q,\,p;\,\beta) &=& \frac{\mathcal{Z}_0(\beta)}{ \mathcal{Z}_{St}(\beta)}\,W_{0}(q,\,p;\,\beta)
\left\{
1 +\hbar^{2}
\left[
\frac{\beta^3}{24}
\left(
\partial_q^{2}V(q)\,\left(\partial_p K(p)\right)^2
+
\partial_p^{2}K(p)\,\left(\partial_q V(q)\right)^2
\right)
\right.\right.
\nonumber\\ &&\qquad\qquad\qquad\qquad\qquad\qquad\qquad\qquad \left.\left.-\frac{\beta^2}{8}
\partial_q^{2}V(q)\,\partial^2_p K(p)
\right]\right\},
\end{eqnarray}
where the multiplying factor, ${\mathcal{Z}_0(\beta)}/{ \mathcal{Z}_{St}(\beta)}$, has been introduced in order to reflect the modifications of the associated partition functions as well as to guarantee the QM unitarity properties.
Of course, for $K(p) \sim p^2$, one recovers the original Wigner results \cite{Wigner,Coffey07}.

To summarize, the same systematic corrections can be used to compute the Wigner currents which, up to order $\mathcal{O}(\hbar^2)$, are written as
\begin{equation}
J^{(2)}_q(q,\,p;\,t) = + \left\{ \partial_p K(p) \left(1 + \hbar^{2} \chi_{(q,\,p;\,\beta)}\right) 
- \frac{\hbar^{2}}{24}\,\partial_p^3 K(p)\,
\left[\beta^2 \left(\partial_q V(q)\right)^2 -\beta \partial^2_q V(q)\right]\right\}W_0,
\label{z500BB}
\end{equation}
and
\begin{equation}
J_p(q,\,p;\,t) = - \left\{ \partial_q V(q) \left(1 + \hbar^{2} \chi_{(q,\,p;\,\beta)}\right) 
- \frac{\hbar^{2}}{24}\,\partial_q^3 V(q)\,
\left[\beta^2 \left(\partial_p K(p)\right)^2 -\beta \partial^2_p K(p)\right]\right\}W_0,
\label{z500BB}
\end{equation}
which can also be recast in the form of
\begin{equation}
J_q(q,\,p;\,t) \approx + \left\{ \partial_p K(p) 
- \frac{\hbar^{2}}{24}\,\partial_p^3 K(p)\,
\left[\beta^2 \left(\partial_q V(q)\right)^2 -\beta \partial^2_q V(q)\right]\right\}W_{St},
\label{z500BB}
\end{equation}
and
\begin{equation}
J^{(2)}_p(q,\,p;\,t) \approx - \left\{ \partial_q V(q) 
- \frac{\hbar^{2}}{24}\,\partial_q^3 V(q)\,
\left[\beta^2 \left(\partial_p K(p)\right)^2 -\beta \partial^2_p K(p)\right]\right\}W_{St}.
\label{z500BB}
\end{equation}

Naturally, according to the truncation criterion, the Wigner perturbed function contributions due to $W_{2\eta}(q,\,p;\,\beta)$, for $\eta$ higher than $1$, could also be evaluated in terms of the derivatives of $K(p)$ and $V(q)$.
Different from Hamiltonian systems that, on the face of the quadratic momentum contributions, are quantum mechanically driven by the non-relativistic Schr\"odinger equation, in the scope of quantum and classical TD, the non-linear effects from $K(p)$ give rise to a non-Gaussian behavior of the momentum distribution, even if the equilibrium phase-space (Maxwell-Boltzmann) distribution function is factorable in the position and momentum variables.
It is independent of higher order quantum correction terms of $W_{St}$, which can be exhaustively computed to any order in powers of $\hbar$.

From now on, aiming to understand the decoupling of quantum corrections from non-linear effects, a more comprehensive view of the phase-space dynamics can be obtained if it is cast in terms of a dimensionless version of $H^{W}(q,\,p)$ from Eq.~\eqref{nlh},
\begin{equation}
\label{dimHH}\mathcal{H}(x,\,k) = \mathcal{K}(k) + \mathcal{V}(x). 
\end{equation}
In this case, $\mathcal{H}(x,\,k)$ is written in terms of dimensionless variables, $x = \left(m\,\omega\,\hbar^{-1}\right)^{1/2} q$ and $k = \left(m\,\omega\,\hbar\right)^{-1/2}p$, such that $\mathcal{H} = (\hbar \omega)^{-1} H$, $\mathcal{V}(x) = (\hbar \omega)^{-1} V\left(\left(m\,\omega\,\hbar^{-1}\right)^{-1/2}x\right)$ and $\mathcal{K}(k) = (\hbar \omega)^{-1} K\left(\left(m\,\omega\,\hbar\right)^{1/2}k\right)$, where $m$ is a mass scale parameter and $\omega$ is an arbitrary angular frequency.
Therefore the Wigner function can be cast into the dimensionless form of $\mathcal{W}(x, \, k;\,\omega t) \equiv \hbar\, W(q,\,p;\,t)$, with $\hbar$ absorbed by $dp\,dq\to \hbar\, dx\,dk$ integrations, i.e.
\begin{eqnarray}\label{alexDimW}
\mathcal{W}(x, \, k;\,\tau) &=& \pi^{-1} \int^{+\infty}_{-\infty} \hspace{-.35cm}dy\,\exp{\left(2\, i \, k \,y\right)}\,\phi(x - y;\,\tau)\,\phi^{\ast}(x + y;\,\tau),
\end{eqnarray}
with $y = \left(m\,\omega\,\hbar^{-1}\right)^{1/2} s$ and $\tau = \omega t$\footnote{In this case, the position and momentum wave functions, $\varphi(x,\,\tau)$ and $\psi(q;\,t)$, are consistently normalized by
\begin{equation}
\int^{+\infty}_{-\infty} \hspace{-.2 cm}{dx}\,\vert\phi(x;\,\tau)\vert^2 =\int^{+\infty}_{-\infty}\hspace{-.2 cm}{dq}\,\vert\psi(q;\,t)\vert^2 = 1.
\end{equation}} 
The Wigner currents are now given by
\begin{eqnarray}\label{alexDimW2222}
\label{imWA}\mathcal{J}_x(x, \, k;\,\tau) &=& +\sum_{\eta=0}^{\infty} \left(\frac{i}{2}\right)^{2\eta}\frac{1}{(2\eta+1)!} \, \left[\partial_k^{2\eta+1}\mathcal{K}(k)\right]\,\partial_x^{2\eta}\mathcal{W}(x, \, k;\,\tau),\\
\label{imWB}\mathcal{J}_k(x, \, k;\,\tau) &=& -\sum_{\eta=0}^{\infty} \left(\frac{i}{2}\right)^{2\eta}\frac{1}{(2\eta+1)!} \, \left[\partial_x^{2\eta+1}\mathcal{V}(x)\right]\,\partial_k^{2\eta}\mathcal{W}(x, \, k;\,\tau),
\end{eqnarray}
so that $\omega\, \partial_x\mathcal{J}_x \equiv \hbar\, \partial_q J_q(q,\,p;\,t)$ and $\omega \,\partial_k\mathcal{J}_k\equiv \hbar \,\partial_p J_p(q,\,p;\,t)$, through which the dimensionless continuity equation is written in terms of the re-defined phase-space coordinates, $\mbox{\boldmath $\xi$} = (x,\,k)$, as
\begin{equation}
{\partial_{\tau} \mathcal{W}} + {\partial_x \mathcal{J}_x}+{\partial_k \mathcal{J}_k} = {\partial_{\tau} \mathcal{W}} + \mbox{\boldmath $\nabla$}_{\xi}\cdot\mbox{\boldmath $\mathcal{J}$} =0.
\end{equation}

Turning back to the TD ensembles, the stationary Wigner function from Eq.~\eqref{oiue} can be written in a dimensionless form,
\begin{eqnarray}\label{oideem}
\mathcal{W}^{(2)}_{St}(x,\,k;\,\beta) &=& \frac{\mathcal{Z}_0(\beta)}{ \mathcal{Z}_{St}(\beta)}\,\mathcal{W}_{0}(x,\,k;\,\beta)
\left\{
1 +
\left[
\frac{(\beta\hbar\omega)^3}{24}
\left(
\partial_x^{2}\mathcal{V}(x)\,\left(\partial_k \mathcal{K}(k)\right)^2
+
\partial_k^{2}\mathcal{K}(k)\,\left(\partial_x \mathcal{V}(x)\right)^2
\right)
\right.\right.
\nonumber\\ &&\qquad\qquad\qquad\qquad\qquad\qquad\qquad\qquad \left.\left.-\frac{(\beta\hbar\omega)^2}{8}
\partial_x^{2}\mathcal{V}(x)\,\partial^2_k \mathcal{K}(k)
\right]\right\},
\end{eqnarray}
in this case, up to order $\mathcal{O}\left((\beta\hbar\omega)^3\right)$.
From Eq.~\eqref{oideem}, it is interesting to note that, for quantum approached TD ensembles, despite the asymptotic stationarity, the non-Liouvillian feature is shown by the non-vanishing value of $\mbox{\boldmath $\nabla$}_{\xi} \cdot \mathbf{w}$, which is written in the dimensionless form
\begin{equation}\label{oideem33}
\omega^{-1}\mbox{\boldmath $\nabla$}_{\xi} \cdot \mathbf{w} = \frac{(\beta\hbar\omega)^2}{12}
\left(
\partial_k^3 \mathcal{K}(k)\partial^2_x \mathcal{V}(x)\partial_x \mathcal{V}(x)
-
\partial_x^3 \mathcal{V}(x)\partial^2_k \mathcal{K}(k)\partial_k \mathcal{K}(k)\right).
\end{equation}
Of course, for the exact quantum description, only evinced in case of knowing the system spectral decomposition, the quantum propagator (Green's function) for a time-independent dimensionless Hamiltonian $\mathcal{H}$ could be expressed by 
\begin{equation}
\Delta(x,t;\,x^{\prime},0) = \langle x \vert \exp(-i\,\tau\, \hat{\mathcal{H}}) \vert x^{\prime} \rangle,
\end{equation}
which, for a canonical ensemble in equilibrium with a heat reservoir at temperature $\mathcal{T}$, corresponds to the inception of an associated thermal density matrix, $\rho(x,\,x^{\prime};\,\beta \hbar \omega)$, obtained \cite{Hillery,Ballentine} by replacing the above related time-dependence, $\tau$, by $- i \, \beta \hbar \omega$.
One thus would have
\begin{equation}
\rho(x,\,x^{\prime};\,\beta \hbar \omega) = \Delta(x,- i \, \beta;\,x^{\prime},0) = \sum_n \exp(-\beta\hbar\omega\, \varepsilon_n)\,\psi^{*}_n({x})\psi_n({x^{\prime}}),
\end{equation}
where the Hamiltonian eigenstates, $\psi_n({x})$, and eigenvalues, $\varepsilon_n$, are implicitly given by $\mathcal{H}\,\psi_n({x}) = \varepsilon_n\,\psi_n({x})$.

In the coordinate representation, the functional $\rho(x,\,x^{\prime};\,\beta \hbar \omega)$ can have its delocalization aspects parameterized by displacement relations, i.e. $x \to x-y$ and $x^{\prime} \to x + y$, so that the $y$-Fourier transform of $\rho(x-y,\,x+y;\,\beta \hbar \omega)$ is identified with a thermalized phase-space probability distribution,
\begin{eqnarray}
\label{W222}
\Omega(x, \, k;\,\beta) &=& \pi^{-1} \int^{+\infty}_{-\infty} \hspace{-.3cm}dy\,\exp{\left(2\, i \, k \,y\right)}\,\rho(x-y,\, x+y;\,\beta\hbar\omega),
\end{eqnarray}
which satisfies the Bloch equation, 
\begin{equation}
\frac{\partial {\Omega}}{\partial \beta} = - \hat{\mathcal{H}}{\Omega} = -{\Omega}\hat{\mathcal{H}}, \qquad {\Omega}(\beta=0)\equiv \mathbb{I}.
\end{equation}
One thus identifies a correspondence between $\rho(x-y,\, x+y;\,\beta\hbar\omega)$ and $\mathcal{Z}^{-1}\exp(-\beta\hbar\omega\, \hat{\mathcal{H}}) \equiv \mathcal{Z}^{-1}{\Omega}$, from which a large set of systematic TD and statistical results follows from the definition of the partition function as a trace-related functional given by
$\mathcal{Z}\equiv\mathcal{Z}(\beta) = Tr[\exp(-\beta\hbar\omega\, \hat{\mathcal{H}})]$.
Therefore, from Eq.~\eqref{W222}, the thermalized (QM) Wigner function can be written as
\begin{eqnarray}
\label{W222B}
\mathcal{W}_\Omega(x, \, k;\,\beta) &=& (\mathcal{Z}\,\pi)^{-1} \int^{+\infty}_{-\infty} \hspace{-.3cm}dy\,\exp{\left(2\, i \, k \,y\right)}\,\rho(x-y,\, x+y;\,\beta),
\end{eqnarray}
normalized by
\begin{equation}\label{stand}
\mathcal{Z}(\beta) = Tr\left[\exp(-\beta\hbar\omega\, \hat{\mathcal{H}})\right] = \int^{+\infty}_{-\infty} \hspace{-.3cm} dx\,\int^{+\infty}_{-\infty} \hspace{-.3cm}dk\,\,\Omega(x, \, k;\,\beta) = \sum_{n=0}^{\infty}\exp\left(-\beta\hbar\omega\,\varepsilon_n\right).
\end{equation}

Finally, as matter of completeness, it is worth mentioning that results from Eqs.~\eqref{oiue}-\eqref{oideem33} can also be interpreted as an extension (from $k^2$ to generic $\mathcal{K}(k)$ Hamiltonian kinetic contributions) of the semiclassical truncated Wigner approximation (TWA). The TWA is shown by noticing that classical trajectories are identified from trajectories along which the Wigner function is conserved \cite{Pol2010,Hillery,Zurek03}, which is indeed also supported by Wigner's seminal proposal \cite{Wigner}. In particular, the TWA is also constrained by the non positive-definite property of the Wigner function such that the corresponding outcome approached function cannot be interpreted as a probability of a particular realization of the boundary conditions \cite{Pol2010}. As pointed out in the broader context of phase-space quantum dynamics \cite{Pol2010}, several slightly different frameworks have been historically considered to circumvent such an issue, even in the context of $k^2$-dependent Hamiltonians.

\section{Gaussian ensembles}

Gaussian ensembles are at the core of information issues in continuous variable QM due to their manipulation properties for building vacuum states, thermal states and coherent quantum states \cite{FidDef01}, as well as for describing atomic ensembles \cite{Michael}.
Besides its theoretical appeal and its experimental relevance in quantum optics and low dimensional physics, the phase-space representation of Gaussian ensembles can also be helpful in establishing the bridge between classical and quantum dynamics.
In fact, Gaussian Wigner functions have been parametrically worked out in order to describe sets of measured data \cite{Niranjan, Bertet} correlated to the issues of classical-quantum correspondence \cite{Giulim}, to probe some hypotheses of quantum chaos \cite{Kowalewska}, and to understand the quantum and classical correlations emerging from non-commutative QM \cite{Bernardini13B,PhysicaA,Bernardini13C,Bernardini13E,Leal2019}.

Given that the WW phase-space QM is closely connected to the information content of a quantum system \cite{Robinett,Georgescu,Adesso,Weedbrook,fide2,Bernardini2021}, the evaluation of decoherence, stationarity and non-classicality aspects arising from the dynamical behavior of Gaussian ensembles might be relevant in describing more accurately the boundaries between classical and quantum regimes.
Considering that the phase-space framework presented in Sec.~II can be applied to a relevant set of non-linear Hamiltonian systems, which includes the Harper Hamiltonian \cite{Harper}, in the following, stationarity and Liouvillianity quantifiers for Gaussian ensembles driven by $H^W(q,\,p)$, Eq.~\eqref{nlh}, will be analytically computed.

One thus considers a Gaussian Wigner function written as
\begin{equation}
G_\gamma(q,\,p) = \frac{\gamma^2}{\pi\hbar}\, \exp\left[-\frac{\gamma^2}{\hbar}\left(\frac{q^2}{a^2}+ a^2\,p^2\right)\right],
\end{equation}
where, in the case of harmonic states, the parameter $a$ is identified with $a = (m\,\omega)^{-1}$, for the mass scale, $m$, and the arbitrary angular frequency, $\omega$, previously introduced.
Again, the problem can be recast into a dimensionless configuration with the Gaussian distribution written as
\begin{equation}
\mathcal{G}_\gamma(x,\,k) = \hbar \,G_\gamma(q,\,p) = \frac{\gamma^2}{\pi}\, \exp\left[-\gamma^2\left(x^2+ k^2\right)\right],
\end{equation}
which leads to the following associated Wigner flow contributions,
\begin{eqnarray}
\label{imWA2}\partial_x\mathcal{J}_x(x, \, k;\,\tau) &=& +\sum_{\eta=0}^{\infty} \left(\frac{i}{2}\right)^{2\eta}\frac{1}{(2\eta+1)!} \, \left[\partial_k^{2\eta+1}\mathcal{K}(k)\right]\,\partial_x^{2\eta+1}\mathcal{G}_{\gamma}(x, \, k),
\\
\label{imWB2}\partial_k\mathcal{J}_k(x, \, k;\,\tau) &=& -\sum_{\eta=0}^{\infty} \left(\frac{i}{2}\right)^{2\eta}\frac{1}{(2\eta+1)!} \, \left[\partial_x^{2\eta+1}\mathcal{V}(x)\right]\,\partial_k^{2\eta+1}\mathcal{G}_{\gamma}(x, \, k),
\end{eqnarray}
once the Hamiltonian form from Eq.~(\ref{dimHH}) has been assumed.
From Gaussian relations with Hermite polynomials of order $n$, $\mbox{\sc{h}}_n$, one has
\begin{equation}
\partial_\zeta^{2\eta+1}\mathcal{G}_{\gamma}(x, \, k) = (-1)^{2\eta+1}\gamma^{2\eta+1}\,\mbox{\sc{h}}_{2\eta+1} (\gamma \zeta)\, \mathcal{G}_{\gamma}(x, \, k),
\end{equation}
for $\zeta = x,\, k$, which can be reintroduced in Eqs.~\eqref{imWA2} and \eqref{imWB2} to lead to potentially convergent series expansions. These allow for recasting the Wigner flow expressions into an analytical form, which accounts for the overall quantum modifications, i.e. for $\eta$ from $1$ to $\infty$ into Eqs.~(\ref{imWA2})-(\ref{imWB2}).
In particular, for the quantum systems where $\mathcal{V}$ and $\mathcal{K}$ derivatives can be recast in the form
\begin{eqnarray}
\label{t111}
\partial_x^{2\eta+1}\mathcal{V}(x) &=& \lambda^{2\eta+1}_{(x)} \, \upsilon(x),\\
\label{t222}
\partial_k^{2\eta+1}\mathcal{K}(k) &=& \mu^{2\eta+1}_{(k)} \, \kappa(k),
\end{eqnarray}
with $\lambda$, $\upsilon$, $\mu$, and $\kappa$ identified as arbitrary auxiliary functions, it can be straightforwardly verified that, once they are substituted into Eqs.~\eqref{imWA2} and \eqref{imWB2}, the above expressions lead to
\begin{eqnarray}
\label{imWA3}\partial_x\mathcal{J}_x(x, \, k;\,\tau) &=& (+2i) \kappa(k)\,\mathcal{G}_{\gamma}(x, \, k)\,\sum_{\eta=0}^{\infty} \left(\frac{i\,\gamma\,\mu_{(k)}}{2}\right)^{2\eta+1}\frac{1}{(2\eta+1)!} \, \mbox{\sc{h}}_{2\eta+1} (\gamma x),\\
\label{imWB3}\partial_k\mathcal{J}_k(x, \, k;\,\tau) &=& (-2i) \upsilon(x)\,\mathcal{G}_{\gamma}(x, \, k)\sum_{\eta=0}^{\infty} \left(\frac{i\,\gamma\, \lambda_{(x)}}{2}\right)^{2\eta+1}\frac{1}{(2\eta+1)!} \, \mbox{\sc{h}}_{2\eta+1} (\gamma k).\end{eqnarray}
Finally, by noticing that
\begin{equation}
\sum_{\eta=0}^{\infty}\mbox{\sc{h}}_{2\eta+1} (\gamma \zeta)\frac{s^{2\eta+1}}{(2\eta+1)!} = \sinh(2s\,\gamma\zeta) \exp[-s^2],
\end{equation}
one gets
\begin{eqnarray}
\label{imWA4}\partial_x\mathcal{J}_x(x, \, k;\,\tau) &=& -2 \kappa(k)\,\sin\left(\gamma^2 \mu_{(k)}\,x\right)\,\exp[+\gamma^2 \mu^2_{(k)}/4]\,\mathcal{G}_{\gamma}(x, \, k)\,
,\\
\label{imWB4}\partial_k\mathcal{J}_k(x, \, k;\,\tau) &=& +2 \upsilon(x)\,\sin\left(\gamma^2 \lambda_{(x)}\,k\right)\,\exp[+\gamma^2 \lambda^2_{(k)}/4]\,\mathcal{G}_{\gamma}(x, \, k),
\end{eqnarray}
which, as prescribed, points to a convergent series for the stationarity quantifier, $\mbox{\boldmath $\nabla$}_{\xi}\cdot \mbox{\boldmath $\mathcal{J}$}$. In addition, depending on the explicit form of the Hamiltonian, it can be manipulated to give the Liouvillian quantifier, $\omega^{-1}\mbox{\boldmath $\nabla$}_{\xi} \cdot \mathbf{w}$ and the complete pattern of the associated Wigner flow.

\section{The Harper-like dynamics}

First introduced by Harper \cite{Harper} for parametrizing the behavior of electrons coupled to magnetic fields in a two-dimensional lattice, the so-called Harper Hamiltonian framework turned into an outstanding tool for several applications \cite{Nat010,Nat020}. These include the description of fractal structures related to the Hofstadter spectral decomposition connected to the phenomenology of the quantum Hall effect \cite{Prange}, and of the quantum mechanical topological phases which emerge in the context of designing ultra-cold atom platforms for producing synthetic gauge fields and topological structures for neutral atoms \cite{001,002}.

From an effective perspective, Harper-like models can be reduced to a one-dimensional Hamiltonian description 
of nearest-neighbor couplings with a senoidal modulation of the on-site energies expressed by a Hamiltonian constraint given by
\begin{equation}\label{HamHarper00}
{H} \psi_n = -A_k (e^{+i\vartheta}\,\psi_{n+1} + e^{-i\vartheta}\,\psi_{n-1}) - A_x\,\cos(2\pi
\alpha\, n +\theta) \psi_n,\end{equation}
where $A_{x}$ and $A_{k}$, correspond to the coupling magnitude and the modulation \cite{PRL-Harper} , respectively, and the phases $\theta$ and $\vartheta$ are related to the wave number in two-dimensions. 

By observing that the displacement of quantum numbers exhibited by $\psi_{n\pm1}$ in Eq. (\ref{HamHarper00}) is associated with localized states in adjacent sites, quantum states described by $\psi_{n\pm1}\sim \psi({x\pm a}) \equiv \exp[\pm i\,k\,a]\psi(q)$ can be parameterized by the action of the translation operators, $\exp[\pm i\,k\,a]$. According to the application of the dimensionless momentum operator, $k$, for a coordinate correspondence given by $(x,\,a)\to(2\pi\alpha n, 2\pi\alpha)$, which is also dimensionless, the Hamiltonian (\ref{HamHarper00}) admits a semi-classical representation reduced to the form of Eq.~\eqref{HamHarper01}.
One can indeed demonstrate that Eq. (\ref{HamHarper01}) is derived from Eq. (\ref{HamHarper00}) \cite{Harper02}, for corresponding coordinate operators, $\hat{x}$ and $\hat{k}$, satisfying $[\hat{x},\hat{k}] = i\,2\pi\alpha$. The so-called Peierls phase parameter, $2\pi\alpha$, plays the role of an effective Planck constant through the condition $2\pi\alpha \equiv 1$, in the dimensionless variables arising from the canonical commutation relation, $[\hat{x},\hat{k}] = i$.
It is also relevant to notice that questions related to quantum chaos are more suitably addressed in case of having $A_{k}=A_{x}=2\pi\alpha \equiv 1$, through which the transition from integrable to chaotic Hamiltonian regimes are more evinced.

\begin{figure}[h]
\includegraphics[scale=0.5]{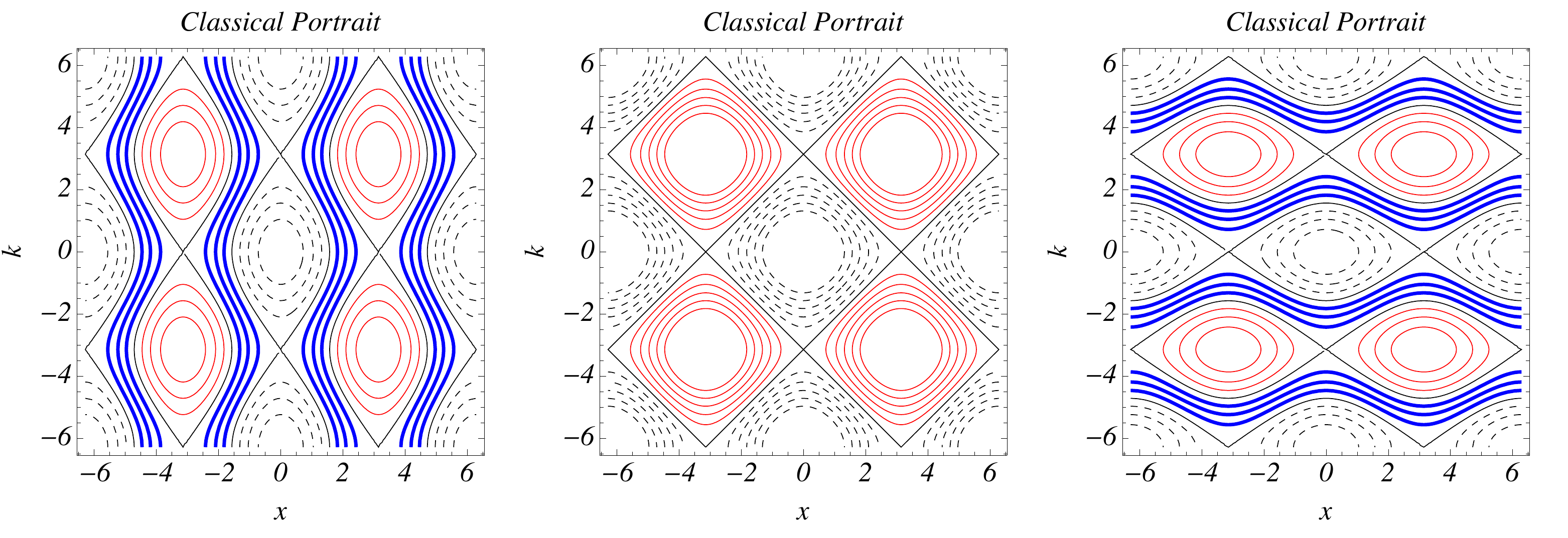}
\renewcommand{\baselinestretch}{.85}
\caption{\footnotesize
(Color online) Classical portrait of Harper Hamiltonians. Phase-space trajectories and corresponding lattice designs are for $ \mbox{Max}\{\nu^2-1,0\} < \vert\epsilon\vert < \nu^2+1$ associated to closed trajectories for $\epsilon > 0$ (black dashed lines) and for $\epsilon < 0$ (red thin lines), and for $0 < \vert\epsilon\vert < \nu^2-1$ corresponding to open trajectories (blue thick lines), when they exist. The threshold (opened-closed) value is given by $\vert\epsilon\vert = \nu^2 -1$. The plots are for $\nu^2=2$ (left), with $\vert\epsilon\vert = 5/2,\, 2,\,3/2,\,\dots,\,0$, $\nu^2=1$ (middle), with $\vert\epsilon\vert = 5/2,\,2,\,3/2,\,\dots,\,0$ and $\nu^2=1/2$ (right), with $\vert\epsilon\vert = 5/2,\, 2,\,3/2,\,\dots,\,0$.}
\label{HarperHarper}
\end{figure}

Of course, the phase-space representation of the electron dynamics in a two-dimensional cristal is also covered by Eq.~\eqref{HamHarper00} which, in order to fit the non-linear properties described by $\mathcal{H}(x,\,k) = \mathcal{K}(k) + \mathcal{V}(x)$, ignoring the global relative sign and admitting a phenomenological variation driven by an arbitrary parameter $\nu^2$, can now be cast into the dimensionless form
\begin{equation}\label{HamHarper01dim}
\mathcal{H}_H(x,\,k)= \cos(k) +\nu^2 \cos(x),
\end{equation}
which exhibits (time-dependent) cyclic analytical solutions. This Hamiltonian works as a feasible platform for identifying classicality and quantumness through the Wigner flow deviations from stationarity and Liouvillian regimes, according to the formalism introduced in the previous sections. The classical properties of the Harper-like Hamiltonian \eqref{HamHarper01dim} are depicted in Fig.~\ref{HarperHarper} where the phase-space trajectories associated with the corresponding lattice designs are identified for several values of the classical energy, $\mathcal{H}_H \to \epsilon$, and for the phenomenological parameter, $\nu$.

\subsection{Results for TD ensembles}

By identifying $\mathcal{K}(k)$ and $\mathcal{V}(x)$ with $\cos(k)$ and $\nu^2\cos(x)$, respectively, as in Eq. \eqref{HamHarper01dim}, one straightforwardly writes the Harper-like associated classical distribution as
\begin{equation}
\mathcal{W}_0(x,\,k;\,\beta_{\hbar\omega}) = (\mathcal{Z}_0(\beta_{\hbar\omega}))^{-1}\,\exp\left[-\beta_{\hbar\omega} \left(\cos(k) +\nu^2 \cos(x)\right)\right],
\end{equation}
with the phase-space domain reduced to the intervals $x \in [-\pi,\,\pi]$ and $k \in [-\pi,\,\pi]$, where $\beta_{\hbar\omega} = \beta{\hbar\omega}$, and from which the partition function can be obtained as
\begin{equation}
\mathcal{Z}_0(\beta_{\hbar\omega}) = 
\int^{+\pi}_{-\pi} \hspace{-.35cm}dx\,\int^{+\pi}_{-\pi} \hspace{-.35cm}dk\,\exp\left[-\beta_{\hbar\omega} \left(\cos(k) +\nu^2 \cos(x)\right)\right] =
4\pi^2\, I_0(\beta_{\hbar\omega})\,I_0(\nu^2\beta_{\hbar\omega}),
\end{equation} 	
where $I_n$ is the modified Bessel function of order $n$.
After the mathematical manipulations which result in Eq.~\eqref{oideem}, one obtains
\begin{eqnarray}\label{oideem2}
\mathcal{W}^{(2)}_{St}(x,\,k;\,\beta_{\hbar\omega}) &=& \frac{\mathcal{Z}_0(\beta_{\hbar\omega})}{ \mathcal{Z}_{St}(\beta_{\hbar\omega})}\,\mathcal{W}_{0}(x,\,k;\,\beta)
\left\{
1 -
\left[
\frac{(\beta_{\hbar\omega})^3}{24}
\left(
\nu^2 \cos(k)\sin^2(x) + \cos(x) \sin^2(k)
\right)
\right.\right.
\nonumber\\ &&\qquad\qquad\qquad\qquad\qquad\qquad\qquad\qquad \left.\left.+\frac{(\beta_{\hbar\omega})^2}{8}
\cos(k)\cos(x)
\right]\right\},
\end{eqnarray}	
with\begin{equation}
\mathcal{Z}_{St}(\beta_{\hbar\omega}) = 4\pi^2\, I_0(\beta_{\hbar\omega})\,I_0(\nu^2\beta_{\hbar\omega}) - \frac{1}{24}(\beta_{\hbar\omega})^2 I_1(\beta_{\hbar\omega})\,I_1(\nu^2\beta_{\hbar\omega}),
\end{equation} 
and with the associated Wigner currents given by
\begin{eqnarray}
\mathcal{J}^{(2)}_x(x, \, k;\,\tau) &=&-\sin(k) \left\{1-\frac{\beta_{\hbar\omega}^3 \nu^2}{24}\left(\nu^2 \cos^2(k) \sin(x)+\sin(k) \cos^2(x)\right)\right.\\&&\qquad\qquad \left.+\frac{1}{24} \left[\beta_{\hbar\omega} \nu^2 \cos(x)+\beta_{\hbar\omega}^2 \left(m^4 \sin^2(x) - 3 \nu^2 \cos(k) \cos(x)\right)\right]\right\}\mathcal{W}_{0},\nonumber\\
\mathcal{J}^{(2)}_k(x, \, k;\,\tau) &=& \nu^2 \sin(x)\left\{1-\frac{\beta_{\hbar\omega}^3}{24} \nu^2 \left(\nu^2 \cos^2(k) \sin(x)+\sin(k) \cos^2(x)\right)\right.\\&&\qquad\qquad \left.+\frac{1}{24} \left[\beta_{\hbar\omega} \cos(k) +\beta_{\hbar\omega}^2 \left(\sin^2(k) - 3 \nu^2 \cos(k) \cos(x)\right)\right]\right\}\mathcal{W}_{0},\nonumber
\end{eqnarray}	
which leads to the Liouvillianity quantifier given by
\begin{equation}
\omega^{-1}\mbox{\boldmath $\nabla$}_{\xi} \cdot \mathbf{w} = \frac{(\beta_{\hbar\omega})^2}{12}
\sin(x)\sin(k)\left(\nu^4 \cos(x)
-\nu^2 \cos(k)\right).
\end{equation}

The partition function of classical and quantum ($\mathcal{O}(\hbar^2)$) stationary ensembles
can be considered when obtaining the TD variables, which include the corresponding quantum purity, $\mathcal{P}_{(\beta_{\hbar\omega})}$ \cite{Bernardini2020}, the (dimensionless) internal energy, $\mathcal{E}_{(\beta_{\hbar\omega})} (\equiv E/(\hbar\omega))$, and the (dimensionless) heat capacity, $\mathcal{C}_{(\beta_{\hbar\omega})}(\equiv C/k_B)$, given respectively by
$$
\mathcal{P}_{(\beta_{\hbar\omega})} = \frac{\mathcal{Z}(2\beta_{\hbar\omega})}{\mathcal{Z}^2(\beta_{\hbar\omega})},\quad
\mathcal{E}_{(\beta_{\hbar\omega})} = -\frac{\partial~}{\partial \beta_{\hbar\omega}}\ln\left[{\mathcal{Z}(\beta_{\hbar\omega})}\right]
,\quad
 \mathcal{C}_{(\beta_{\hbar\omega})} = \beta^2_{\hbar\omega}\left(\frac{\partial~}{\partial \beta_{\hbar\omega}}\right)^2\ln\left[{\mathcal{Z}(\beta_{\hbar\omega})}\right].$$

\begin{figure}
(a)\hspace{-1.5 cm}\includegraphics[scale=0.43]{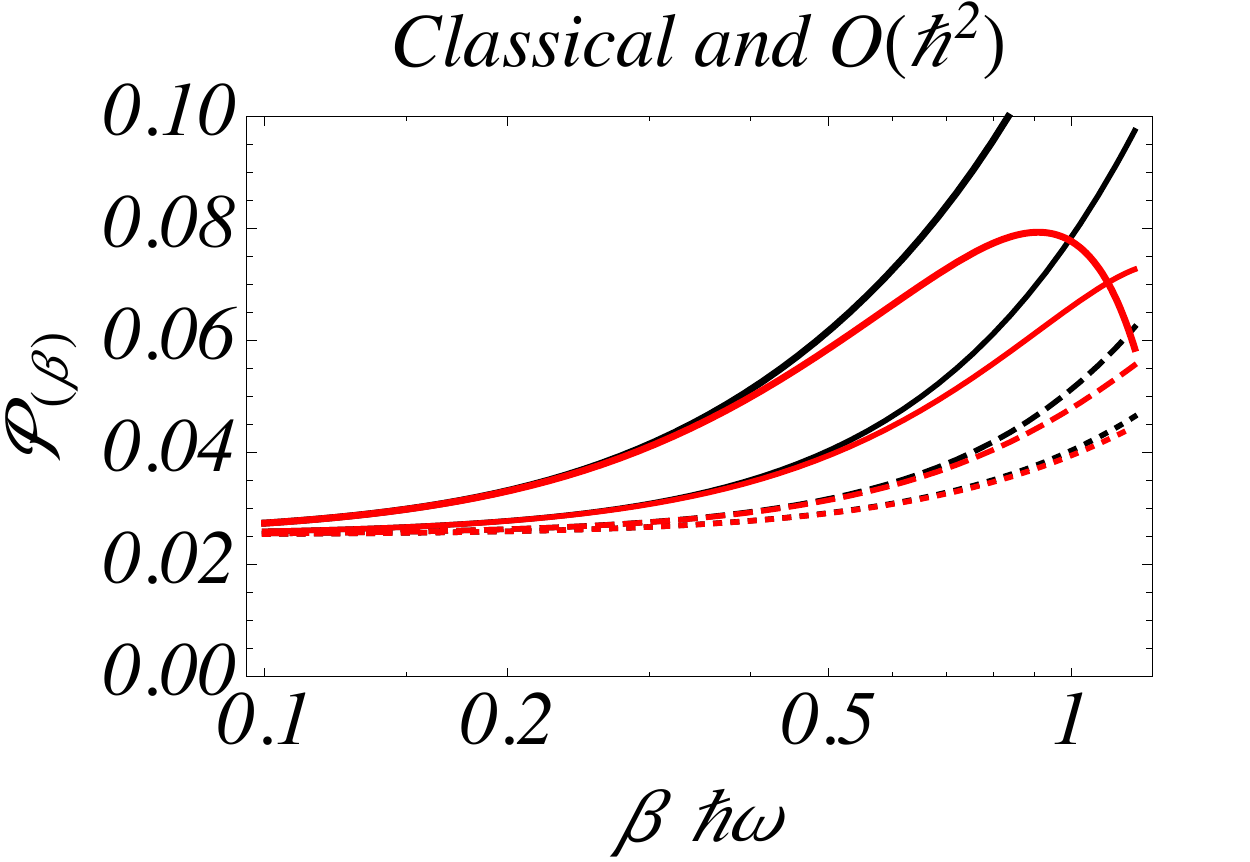}\hspace{1 cm}
(b)\hspace{-1.5 cm}\includegraphics[scale=0.43]{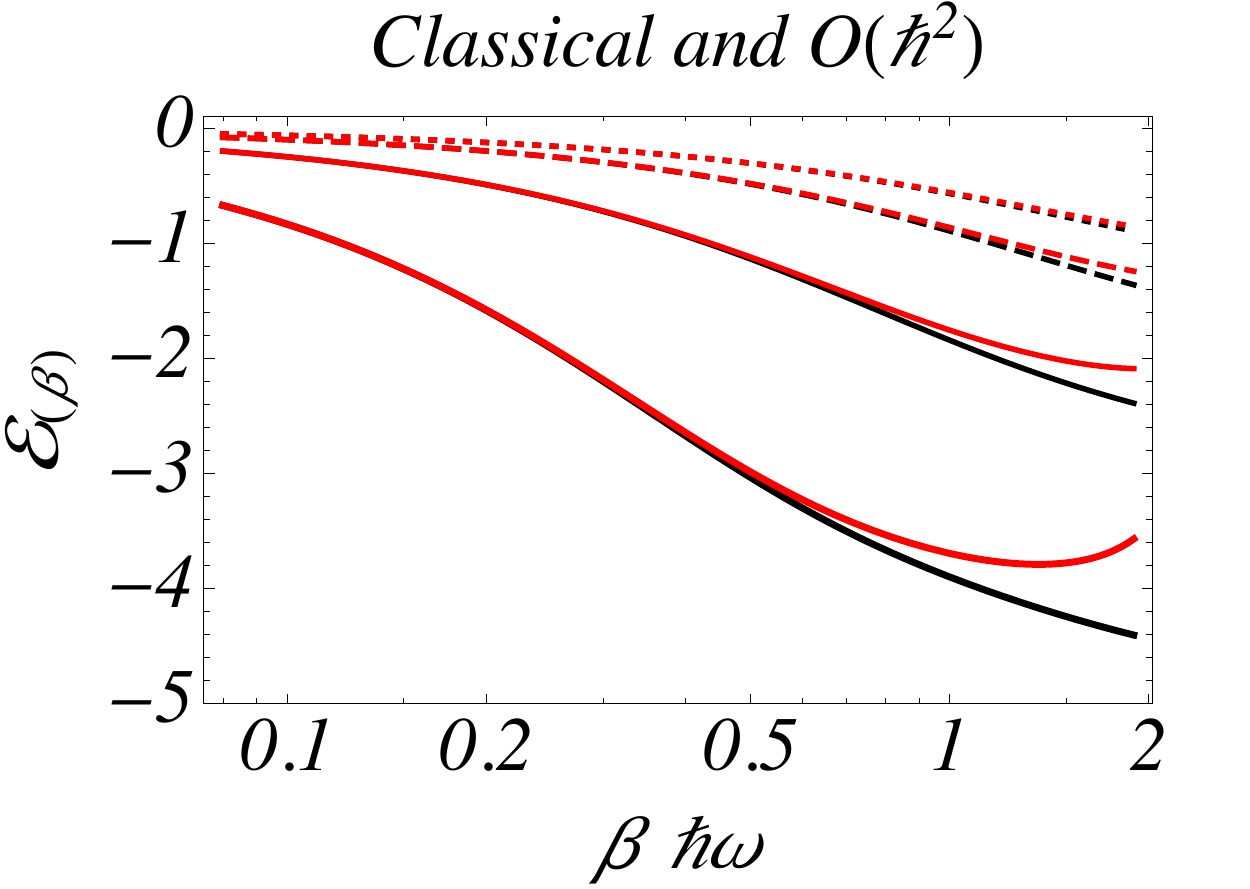}\hspace{1 cm}
(c)\hspace{-1.5 cm}\includegraphics[scale=0.43]{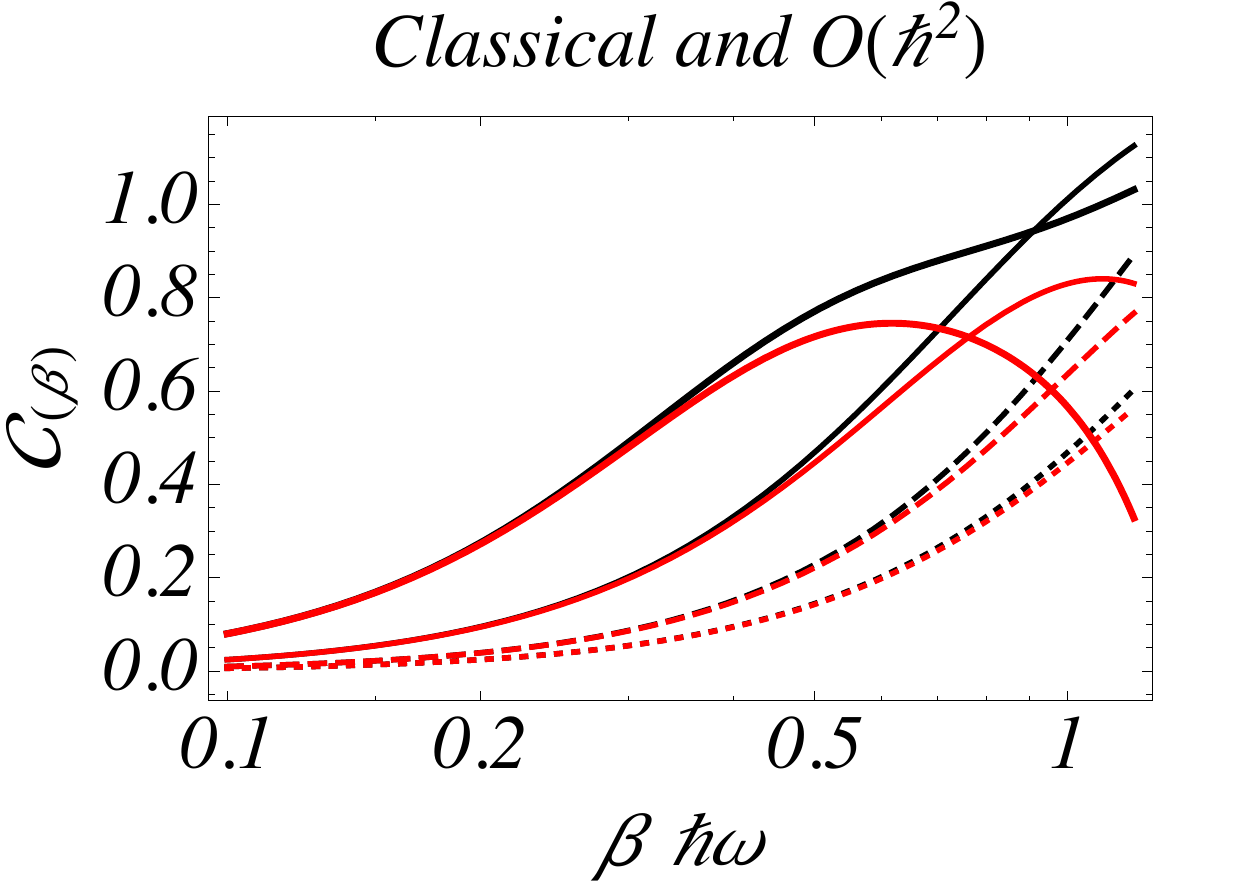}\hspace{1 cm}
\renewcommand{\baselinestretch}{.6}
\caption{\footnotesize
(Color online) (a) Purity (first plot), $\mathcal{P}(\beta_{\hbar\omega})$, (b) Internal energy (second plot), $\mathcal{E}(\beta_{\hbar\omega})$, and (c) Heat capacity (third plot), $\mathcal{C}(\beta_{\hbar\omega})$, as function of $\beta_{\hbar\omega}$, for classical (black lines) and quantum ( $\mathcal{O}(\hbar^2)$, red lines) stationary regimes.
The results are for $\nu^2 = 1/\sqrt{2}$ (dotted lines), $1$ (dashed lines), $\sqrt{2}$ (thin lines) and $2$ (thick lines).}
\label{HarperHarper02}
\end{figure}

From results depicted in Fig.~\ref{HarperHarper02}, one notices that quantum contributions recover the meaning of quantum purity and work to constrain its maximum values to be lower than unity (pure state). The results also exhibit corrections to $\mathcal{E}$ and $\mathcal{C}$ which work fine only for $ 0 < \beta \hbar \omega \lesssim 1$, in an interval that is unaffected by the choice of the Peierls phase parameter, $2\pi\alpha$, set equal to unity.

\begin{figure}
\vspace{-1.2 cm}\includegraphics[scale=0.185]{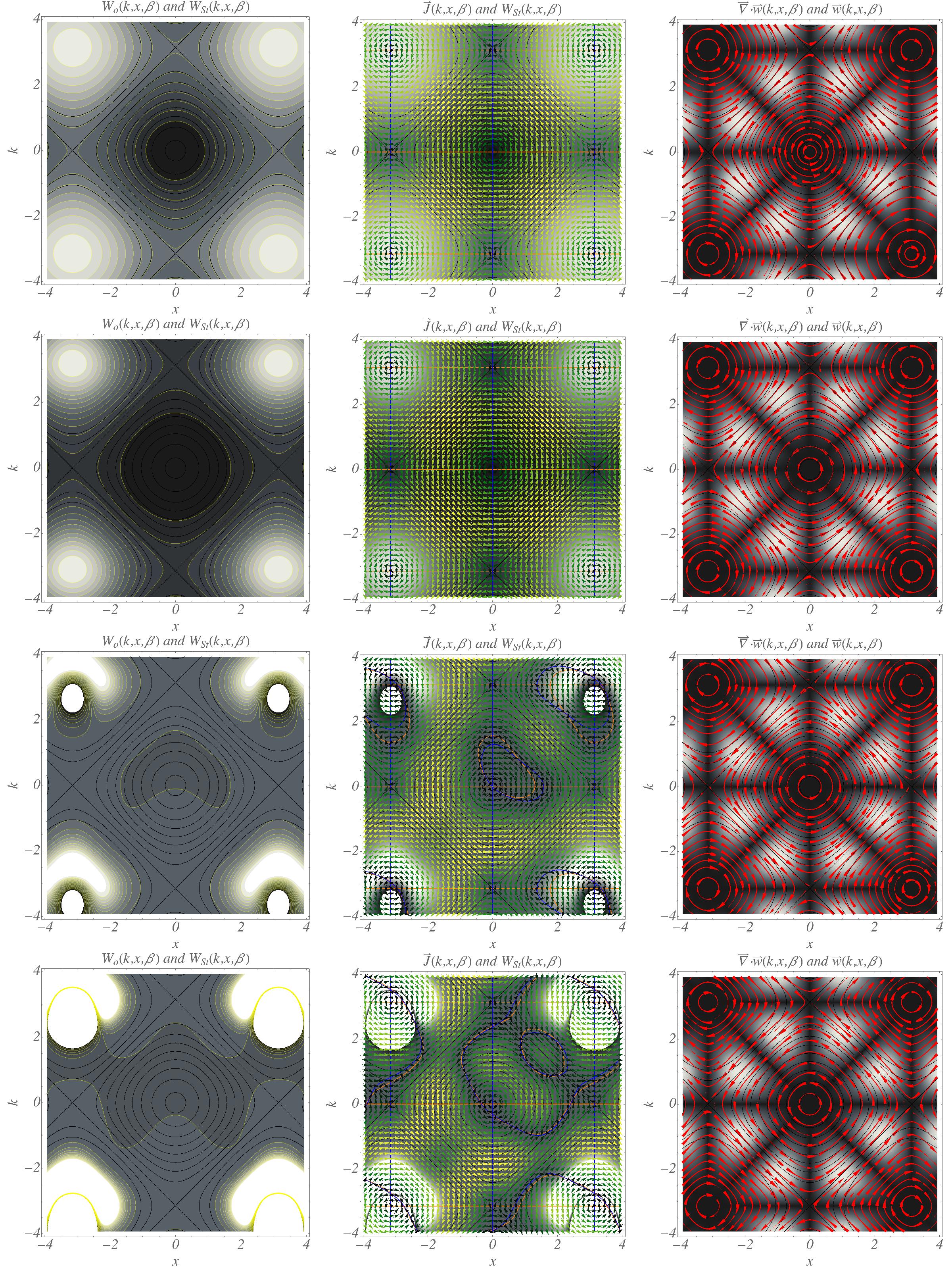}
\renewcommand{\baselinestretch}{.6}
\caption{\footnotesize
(Color online)
{\em First column}: Classical (black contours) and QM corrected (yellow contours) Wigner function profiles, $\mathcal{W}_0(x,\,k;\, \beta_{\hbar\omega})$ and $\mathcal{W}^{(2)}_{St}(x,\,k;\, \beta_{\hbar\omega})$.
{\em Second column}: Features of the Wigner flow for the TD ensemble in the $x - k$ plane. Blue contour lines are for $J^{(2)}_x(x,\,k;\, \beta_{\hbar\omega}) = 0$ and orange contour lines are for $J^{(2)}_k(x,\,k;\, \beta_{\hbar\omega}) = 0$.
The {\em greenyellow} vector arrow color scheme shows the (quantum $\mathcal{O}(\hbar^2)$) Wigner current profiles with the domains of quantum fluctuations bounded by blue and orange lines. Darker green arrows correspond to intenser fluxes. 
{\em Third column}: Liouvillian quantifier, $\mbox{\boldmath $\nabla$}_{\xi} \cdot \mathbf{w}$, superposed by the normalized quantum velocity field representation (red arrows), $\mathbf{w} /\vert\mathbf{w}\vert$. The background color scheme (from darker regions, $\mbox{\boldmath $\nabla$}_{\xi} \cdot \mathbf{w} \sim 0$, to lighter regions, $\mbox{\boldmath $\nabla$}_{\xi} \cdot \mathbf{w} > 0$) reinforces the approximated Liouvillian behavior for the central region.
The results are for $\nu^2 =1$ and $\beta_{\hbar \omega} =0.1$ (first row), $1$ (second row), $3$ (third row) and $5$ (forth row).}
\label{HarperHarper03}
\end{figure}

Besides the natural interpretation of the quantum contributions which lead to an overall suppression of classical values, it is evident that negative values of $\mathcal{E}$ have no classical analogs in the Schr\"odinger theory for Hamiltonians with quadratic kinetic terms.
The internal energy of a real neutral gas depends on how the temperature, $\mathcal{T}$, appears in the equation of state.
In general, negative values for $\mathcal{E}$ means that interaction contributions dominate over thermal kinetic ones, which decrease as $\beta$ increases.
In the same sense, despite evident modifications for $\mathcal{C}$, a more accurate quantum behavior could only be described by accounting for higher order terms in the $\hbar^2$ expansion.
One notices that the results are not qualitatively modified by the parameter $\nu$, which is set equal to $1$ for the evaluation of the Wigner current maps (cf. Fig.~\ref{HarperHarper03}).

To complete the TD ensemble analysis, the non-Liouvillian behavior is depicted in Fig.~\ref{HarperHarper03}. 
The results are reduced to the central site, but they can be periodically reproduced from $x,\,k \in (-\pi,\,\pi)$ to $x,\,k \in (-\pi\pm 2n\pi,\,\pi\pm 2n\pi)$ (with $n \in \mathbb{Z}$) due to the lattice symmetry properties.
For $\beta_{\hbar \omega} \ll 1$ (first row), quantum fluctuations are highly suppressed by the thermal fluctuations and cannot be identified through the pattern of the Wigner function (first column) and of the corresponding Wigner flow (second column), which is depicted by the light-dark background color in Fig.~\ref{HarperHarper03}.
The Wigner flow stagnation points are identified by orange-blue crossing lines, where $\mathcal{J}^{(2)}_x = \mathcal{J}^{(2)}_k=0$, and can be interpreted as a consequence of the $\mathcal{O}(\hbar^2)$ quantum contributions which are evinced for $\beta_{\hbar \omega} \gtrsim 1$. Blue (orange) contour lines are bound for the reversal of the Wigner flow in the $x(k)$ direction.
The quantum fluctuations are identified by clockwise and anti-clockwise vortices (winding number equals to $+1$ and $-1$), separatrix intersections and saddle points (winding number equals to $0$).
The contra-flux fringes (bounded by blue/orange lines) emerge to compensate the retarded evolution of the quantum flux. 
The classical profile does not exhibit such an overall locally compensation phenomena.
For comparison reasons, the classical trajectory is shown as a collection of black thin lines, which do not exhibit any local features like the above-mentioned ones.
The non-Liouvillian quantifier is depicted in the third column of Fig.~\ref{HarperHarper03}, in correspondence with the quantum TD ensemble Wigner function, for $\beta_{\hbar \omega} =0.1$ (first row), $1$ (second row), $3$ (third row) and $5$ (forth row), from which one notices that the quantum effects are much more evident for lower temperatures (increasing $\beta_{\hbar\omega}$ values).

\subsection{Results for Gaussian ensembles}

For Gaussian ensembles driven by $\mathcal{K}(k)=\cos(k)$ and $\mathcal{V}(x)=\nu^2\cos(x)$, exact analytical results can be obtained. 
Firstly, one identifies $\lambda^{2\eta+1}_{(x)} = \mu^{2\eta+1}_{(k)}= (-1)^{\eta +1}$ from Eqs.~\eqref{t111} and \eqref{t222}, which leads to $\mu=\lambda = i$, $\upsilon(x) = i\,\nu^2\,\sin(x)$, and $\kappa(k) = i\,\sin(k)$.
Once they are replaced into Eqs.~\eqref{imWA4} and \eqref{imWB4}, one can write
\begin{eqnarray}
\label{imWA4CC}\partial_x\mathcal{J}_x(x, \, k;\,\tau) &=& +2 \,\sin\left(k\right)\,\sinh\left(\gamma^2\,x\right)\,\exp[-\gamma^2 /4]\,\mathcal{G}_{\gamma}(x, \, k)\,
,\\
\label{imWB4CC}\partial_k\mathcal{J}_k(x, \, k;\,\tau) &=& -2 \nu^2\,\sin\left(x\right)\,\sinh\left(\gamma^2\,k\right)\,\exp[-\gamma^2 /4]\,\mathcal{G}_{\gamma}(x, \, k),
\end{eqnarray}
as a result from the convergent series expansion Eqs.~\eqref{imWA3} and \eqref{imWB3}. 
This means that a Gaussian ensemble driven by the Harper-like Hamiltonian can be considered in order to compare classical and quantum regimes, providing the exact analytic expression for the quantum scenario in the phase-space. 
The integrated Wigner currents obtained from Eqs.~\eqref{imWA4} and \eqref{imWB4} can thus be read as
\begin{eqnarray}
\label{imWA4CCD}\mathcal{J}_x(x, \, k;\,\tau) &=& +\frac{\gamma}{2\sqrt{\pi}} \,\sin\left(k\right)\,\exp\left(-\gamma^2\,k^2\right)\,
\left[\mbox{\sc{Erf}}\left(\gamma(x-1/2)\right)-\mbox{\sc{Erf}}\left(\gamma(x+1/2)\right)\right],\\
\label{imWB4CCD}\mathcal{J}_k(x, \, k;\,\tau) &=& -\frac{\gamma}{2\sqrt{\pi}} \nu^2\,\sin\left(x\right)\,\exp\left(-\gamma^2\,x^2\right)\,
\left[\mbox{\sc{Erf}}\left(\gamma(k-1/2)\right)-\mbox{\sc{Erf}}\left(\gamma(k+1/2)\right)\right],\qquad
\end{eqnarray}
written in terms of error functions, $\mbox{\sc{Erf}}(\dots)$, and of course, the components of the quantum-like velocity, $\mathbf{w}$, $w_x$ and $w_k$, are obtained by replacing $-k^2 \leftrightarrow +x^2$ in the exponential function of the respective Eqs.~\eqref{imWA4CCD} and \eqref{imWB4CCD} (multiplied by $\pi\gamma^{-2}$).

\begin{figure}
\vspace{-1.2 cm}\includegraphics[scale=0.23]{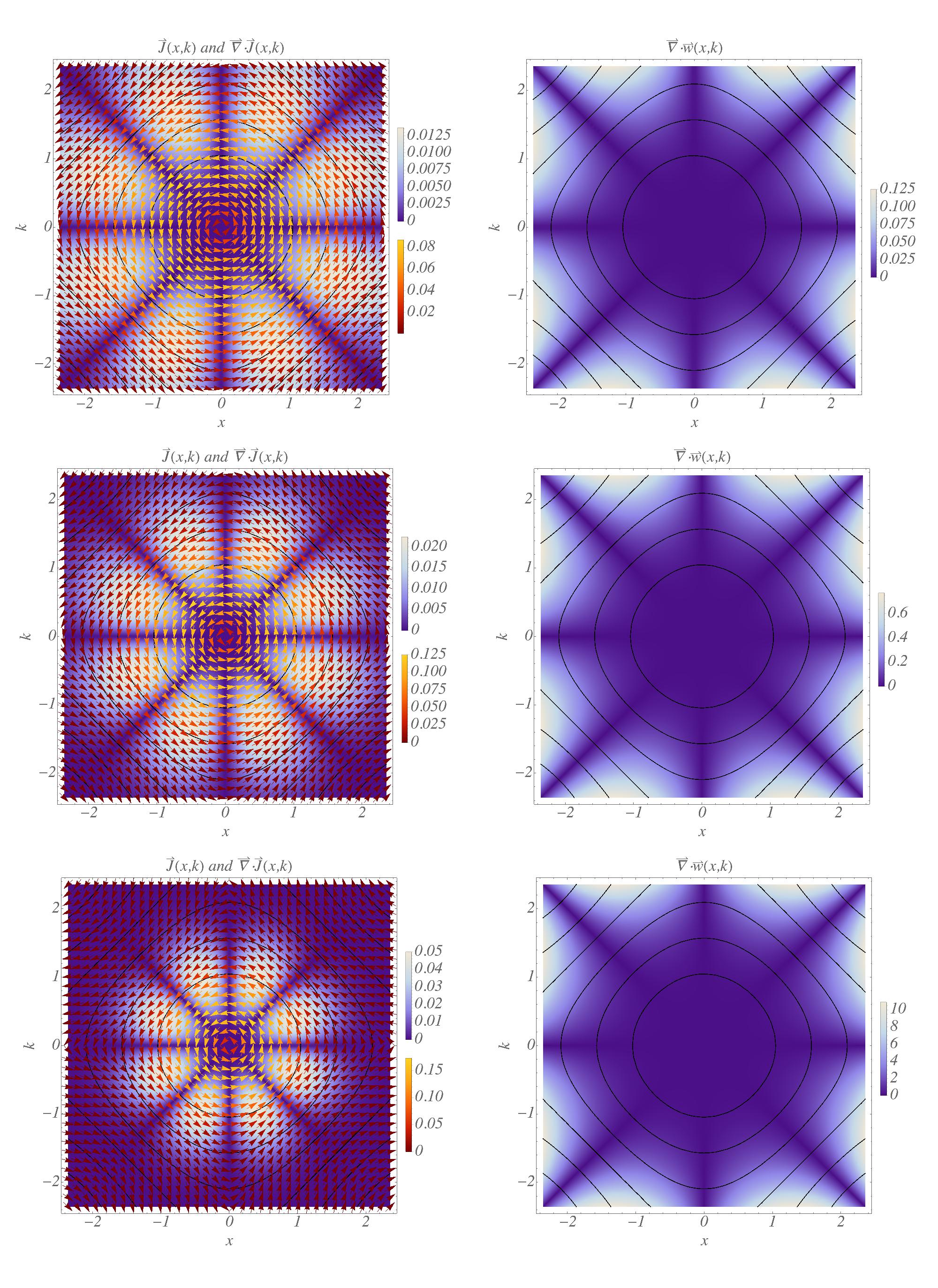}
\renewcommand{\baselinestretch}{.6}
\vspace{-1.2 cm}\caption{\footnotesize
(Color online)
{\em First column}: Features of the Wigner flow for the Gaussian ensemble, in the $x - k$ plane. At $\tau = 0$, Gaussian ensembles do not exhibit neither vortices nor stagnation points, in a kind of camouflage of the quantum corrections.
The stationarity quantifier, $\mbox{\boldmath $\nabla$}_{\xi} \cdot \mbox{\boldmath $\mathcal{J}$}$, the modulus of $\mbox{\boldmath $\mathcal{J}$}$ are described respectively according to the {\em light-dark blue background} and {\em yellowred background} color schemes. The results are for the increasing spreading characteristic of the Gaussian function, from $\gamma =1/\sqrt{2}$ (first row), $1$ (second row) and $\sqrt{2}$ (third row). Peaked Gaussian distributions ($\gamma =\sqrt{2}$) localizes the quantum distortions which result in non-stationarity.{\em Second column}: Liouvillian quantifier, $\omega^{-1}\mbox{\boldmath $\nabla$}_{\xi} \cdot \mathbf{w}$, depicted through the light-dark blue background color scheme, from darker regions, $\mbox{\boldmath $\nabla$}_{\xi} \cdot \mathbf{w} \sim 0$, to lighter regions, $\mbox{\boldmath $\nabla$}_{\xi} \cdot \mathbf{w} > 0$. 
In both columns, classical patterns are shown as a collection of black lines.}
\label{HarperHarper04}
\end{figure}

The departure configuration ($\tau = 0$) of the Gaussian Wigner flow pattern described above is depicted in Fig.~\ref{HarperHarper04}, where the density plot for stationarity and Liouvillianity quantifiers, $\mbox{\boldmath $\nabla$}_{\xi} \cdot \mbox{\boldmath $\mathcal{J}$}$ and $\omega^{-1}\mbox{\boldmath $\nabla$}_{\xi} \cdot \mathbf{w}$, are identified.
	
The stationarity quantifier, $\mbox{\boldmath $\nabla$}_{\xi} \cdot \mbox{\boldmath $\mathcal{J}$}$, is depicted according to the background color scheme, from which lighter regions correspond to non-vanishing local contributions to $\partial_t \mathcal{G}_{\gamma}(x,\,k)$. In spite of the $x-k$ symmetry provided by the Gaussian distribution, one should notice that stationarity and Liouvillianity are decoupled from each other as their associated quantifiers do not exhibit the same local pattern.

Finally, once the above quantifiers, $\mbox{\boldmath $\nabla$}_{\xi} \cdot \mbox{\boldmath $\mathcal{J}$}$ and $\omega^{-1}\mbox{\boldmath $\nabla$}_{\xi} \cdot \mathbf{w}$, are integrated over the volume enclosed by the classical path, $\mathcal{C}$, the quantum effects are shown to be averaged out. It shows that, despite providing the phase-space local identification of quantum effects, Gaussian ensembles, in the case of a Harper-Hamiltonian system, suppress the local vortices from the quantum Wigner pattern, and keep the flow of probability and information properties quantitatively equivalent to those of the classical pattern.

\section{Conclusions}

The phase-space framework for describing the classical-quantum behavior of systems driven by one-dimensional non-linear Hamiltonians described by $H^{W}(q,\,p) = K(p) + V(q)$ (constrained to $\partial ^2 H^{W} / \partial q \partial p = 0$) was introduced and analytically scrutinized through their Wigner flow properties.
Generalized Liouvillian and stationary properties were established for TD and Gaussian quantum ensembles to quantify the quantum modifications over the classical phase-space Hamiltonian solutions.
For TD ensembles, a natural extension of the Wigner approach for perturbatively constructing stationary Wigner functions was provided, to include the Wigner procedure as a particular solution related to the Hamiltonian kinetic contribution identified by $K(p) = p^2/2m$.
Following a parallel procedure, for Gaussian ensembles, the overall quantum distortion over a classical phase-space trajectory was obtained in terms of convergent infinite series expansions over $\hbar^{2}$.

As noticed from previous works, the WW framework, now in its generalized version, can work as a probe for quantumness and classicality for an enlarged class of Hamiltonian systems which includes out-of-the-ordinary contributions for the kinetic term, $K(p) \neq p^2/2m$.
As an implementation example, our results were specialized to a Harper-like Hamiltonian system which, besides working as an experimental platform first identified in Harper's approach for crystal electrons in the presence of magnetic fields \cite{Harper}, admits a statistical description in terms of TD and Gaussian ensembles, like what was here considered.
However, our extended Wigner phase-space approach is not restricted to Harper-like Hamiltonians.
In fact, the Hamiltonian from Eq.~(\ref{HamHarper01}) can be interpreted as a restrictive choice of the phenomenological parameters of the more general Aubry-Andr\'e model \cite{Harper02,PRA-Harper19}, which includes additional linear contributions to a re-parameterized form of $H^{W}(q,\,p)$, with $K(p)\to K(p) + \omega_p p$ and $V(q) \to V(q) + \omega_q q$, where momentum and position contribution coefficients, $\omega_p(p)$ and $\omega_q (q)$ (cf. Eq.~\eqref{HamHarper01}) are the drivers of either bound-state or continuous spectrum regimes.

Beyond the condensed matter physics examples, deformed Hamiltonians with $K(p) = \cosh(p/p_0)$, which are sometimes related to the Toda lattice theory \cite{sigma}, have been conjectured in the investigation of the Seiberg-Witten curve of a $N = 2$ Yang-Mills theory, where a quantization hypothesis follows from the correspondence between spectral theory and strings \cite{Pasquier}.
Following a similar prescription where the hyperbolic behavior is dominant in the description of the competitive ecological equilibrium of populations, Lotka-Volterra-like systems \cite{LV}, besides being extensively applied stochastic systems \cite{Allen,Grasman}, can be treated via the dynamical equations arising from the Hamiltonian formulation described (in a dimensionless notation) by 
\begin{equation}\label{Ham}
\mathcal{H}(x,\,k) = x + k + e^{-x} + e^{-k},
\end{equation}
which yields the following classical equations of motion,
\begin{eqnarray}\label{Ham2}
d{x}/d\tau &=& \{x,\mathcal{H}\}_{PB} = 1-e^{-k},\\
d{p}/d\tau &=& \{k,\mathcal{H}\}_{PB} = e^{-x} - 1,
\end{eqnarray}
whose phase-space trajectories can be extended to the non-commutative context, $[x,\,k]\neq 0$, through the WW formalism. This allows for quantifying (cf. Refs. \cite{NossoPaper,Meu2018}) related phase-space correlations and information flow aspects as well as quantum-like deviations from the well-known prey-predator classical system.

Finally, it is worth mentioning that the procedure discussed here can be equally specialized to the above-mentioned scenarios, and could encompass more complex forms of non-linear Hamiltonians. Clearly, an extended framework of phase-space QM that deserves further research.

{\em Acknowledgments -- The work of A.E.B. is supported by the Brazilian Agencies FAPESP (Grant No. 2020/01976-5) and CNPq (Grant No. 301000/2019-0).}

\end{document}